\documentclass[%
 reprint,
 amsmath,amssymb,
pra,
]{revtex4-2}
\usepackage{booktabs}
\usepackage{multirow}
\usepackage{graphicx}
\usepackage{dcolumn}
\usepackage{bm}

\begin{document}

\preprint{APS/123-QED}

\title{Quantum secure direct communication based on fully passive source  }
\author{Jia-Wei Ying,$^{1}$ Qi Zhang,$^{2}$ Shi-Pu Gu,$^{1}$ Xing-Fu Wang,$^{2}$ Lan Zhou,$^{2}$}\email{zhoul@njupt.edu.cn}
\author{Yu-Bo Sheng$^{1}$}
 \email{shengyb@njupt.edu.cn}
\affiliation{%
 $^1$College of Electronic and Optical Engineering and College of Flexible Electronics (Future Technology), Nanjing
 University of Posts and Telecommunications, Nanjing, 210023, China\\
 $^2$College of Science, Nanjing University of Posts and Telecommunications, Nanjing, 210023, China\\
}%

\date{\today}

\begin{abstract}
 In practical quantum communication, imperfect devices may introduce side channels, creating opportunities for eavesdroppers. Especially on the source side, the side channels created by active modulation may compromise the security of the protocol. We proposes a passively-sourced quantum secure direct communication (QSDC) protocol based on fully passive source.
 By passively modulating both the quantum state and the intensity of the decoy state, 
 we can avoid active modulation operations at the source, thereby enhancing the robustness of QSDC against side-channel attacks. We developed a system model and conducted parameter optimization to obtain the maximum secrecy message transmission rate achievable by the protocol for each channel attenuation.
  At a channel attenuation of 2, 4, 6 dB (corresponding to a communication distance of 5, 10, 15 km), the  secrecy message transmission rates  are $5.76 \times 10^{-5}$($I=0.0895, \bigtriangleup x=0.0490 \pi, \bigtriangleup z=0.0546  \pi$),  $9.92 \times10^{-6}$($I=0.0471, \bigtriangleup x=0.0367 \pi, \bigtriangleup z=0.0408 \pi$), and $4.99\times10^{-7} $($I=0.0168, \bigtriangleup x=0.0152 \pi, \bigtriangleup z=0.0214 \pi$) bit/sec. And its maximum communication distance is about $16.875 $ km, which is about $94.4\%$ of that of actively modulated QSDC.

\end{abstract}

\maketitle


\section{Introduction}

Quantum communication has garnered significant attention due to its unconditional security. Quantum secure direct communication (QSDC) is a branch of quantum communication, which can directly transmit message. The first QSDC protocol was proposed in 2002 \cite{QSDC1},  and subsequent research has yielded numerous theoretical \cite{QSDC2,QSDC3,QSDChd,li,Masking,QSDCos,QSDCdense,QSDChy,QSDC22,QSDC23,QSDCpath,RDIQSDC} and experimental \cite{QSDC9,QSDC10,QSDC11,QSDC12,QSDC15,decoy3,Mapping,QSDC20n,QSDCpath24,cvrecent,QSDCnet,cvexp} results.
As theoretical and experimental understandings deepen, the practical security of quantum communication has become an area of increasing focus \cite{qkd3,QKDsp,PQKD1,collective2,PQSDC1,decoy4}.
Quantum communication is theoretically unconditionally secure, while imperfections in actual experimental equipment can introduce side channels that may be exploited by eavesdroppers \cite{QHACK1,QHACK2,QHACK3,QHACK4,QHACK5}.    The sources and detectors are particularly critical in this context.

In 2012, the measurement-device-independent (MDI) quantum key distribution (QKD) protocol was proposed \cite{qkd2}, which can eliminate all side channels on the detector side, significantly enhancing the security of QKD \cite{MDIQKD1,MDIQKD2,MDIQKD3,MDIQKD4}.
 In the field of QSDC, MDI was applied to QSDC in 2020 and successfully eliminated the side channels associated with QSDC detectors \cite{QSDC14n,MDI2}. Subsequently, there are many studies on MDI-QSDC \cite{MDI3,QSDC16n,QSDC19,QSDC24}.
While these efforts focused on defending against side channels of the detectors, addressing vulnerabilities at the source is equally important.

An effective countermeasure for the imperfect  source is the device-independent(DI) protocol \cite{DI,qkd1,DIQKD,DIQKD1,QSDC16,QSDC18,QSDC21}, which can resist all attacks targeting both detectors and sources.     DI QKD was proposed in 2007 \cite{qkd1} and  experimentally demonstrated in 2022 \cite{DIE1,DIE2,DIE3}.
However, DI is very sensitive to loss, it needs  state-of-the-art devices, and the communication distance and key rate are still limited \cite{DIreview}.

Another measure to improve the security of  sources is to use passive sources.  In quantum communication, we usually need to perform some modulation operations on the source.  This is especially true in the decoy state method \cite{decoy,decoy1,decoy3,decoy4}, where we have to modulate both the light intensity and the quantum state.  These modulation processes, both intensity and state, will introduce side channels, creating opportunities for eavesdropper \cite{qkd3,sidechannel}.
One notable threat is the trojan horse attack \cite{trojan0,trojan1,trojan2,trojan3}. The eavesdropper injects a pulse into the light source, and the trojan horse pulse is modulated with the signal pulse and reflected through the device into the channel. Eve can analyze reflected pulses to obtain information.
The advantage of a passive source is that it eliminates the need for active modulation of the source, thereby removing the side channels associated with the source modulator.  This setup can effectively resist side-channel attacks targeting the modulator.  Passive decoy states \cite{passivedecoy1,passivedecoy2,passivedecoy3,passivedecoy4,passivedecoy5,passivedecoy6,passivedecoy}   and passive coding \cite{passivestate1,passivestatecv1,passivestatecv2,passivestatecv3,QSDCpc} have been widely applied in quantum communication. Recently, the fully passive QKD has been proposed \cite{passive3}, which combines passive decoy states and passive coding.  It employs four phase-random coherent pulses to generate a random quantum state with random intensity through beam splitter(BS) interference and polarization beam splitter(PBS) combination.  By post-selecting the output  photon pulses, the desired state can be obtained.

Passive sources have been incorporated into many quantum communication protocols\cite{passive5,passive6,passive7,passive8,FPTW,FPCKA} and have been experimentally validated\cite{passive9,passive4}.
Inspired by the previous work \cite{passive3,passive5}, we  proposed  a  QSDC protocol based on fully passive source. By integrating fully passive  source with QSDC, we can effectively resist the side channel attack against QSDC  source, so as to 
reduce the demand for the source-side equipment.
This paper mainly makes the following contributions: (1) Based on the fully passive source, we establish a QSDC system model, and give the formula of its secrecy message capacity, as well as the complete analysis process.
(2) We optimize the intensities and the size of the post-selection intervals of the source, 
 and obtain the optimal secrecy message transmission rate corresponding to each channel attenuation.

The structure of the paper is as follows. In Sec. II, we introduce  the passively-sourced QSDC protocol. In Sec. III, we give the secrecy message capacity of the passively-sourced QSDC. In Sec. IV, we establish a system model for the our protocol. In Sec. V, we give the decoy state method. In Sec. VI,  we conduct numerical simulation and parameter optimization. In Sec. VII, we give a  conclusion.

\section{The passively-sourced QSDC protocol}

\begin{figure}[!htbp]
	\begin{center}
		\includegraphics[width=8.5cm,angle=0]{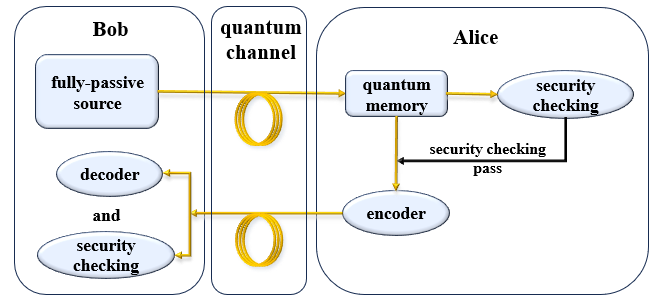}
		\caption{The structure of the passively-sourced QSDC protocol.  }\label{pro}
	\end{center}
\end{figure}

In this section, we explain the passively-sourced QSDC protocol. As shown in  Fig. \ref{pro}, there are two participants, the message sender Alice and the message receiver Bob. The passively-sourced QSDC protocol can be described as follows.

Step 1: Initial state preparation.  The message receiver Bob generates a series of weak coherent pulses through a fully passive source. The intensities of these pulses are random, and the photons in them are in the quantum state $ a^{\dagger } (\theta , \phi )$.
Bob filters the required quantum states and pulse intensities through a post-selection module, classifies them into $X$ basis, $Y$ basis, and $Z$ basis photons according to the quantum states, divides them into $signal$ states, $decoy1$ states, and $decoy2$ states based on the pulse intensities, and then sends these pulses to Alice.
Details about the fully passive source and post-selection intervals will be described  in the section \ref{fpsource}.

Step 2: First round of security checking.  Alice receives these pulses and stores them in quantum memory. Then randomly select a part of the pulses to perform the first round of security checking.
Specifically, Alice will randomly choose the $X$, $Y$, and $Z$ bases to measure the photons. 
Through security  checking, Alice can estimate the error rate of the first-round transmission $E1$.
If $E1$ is below the tolerate threshold, the communication continues. Otherwise, the parties have to discard the communication.

Step 3: Encoding. After the security checking is passed, Alice takes out the remaining photons in the quantum memory, selects some of them for random encoding for the second round of security checking, and encodes the rest for information with the operations $M_{0}$ and $M_{1}$, where  $M_{0}=|H\rangle\langle H|+|V\rangle\langle V|$ and $M_{1}=|H\rangle\langle V|-|V\rangle\langle H|$. $M_{0}$ and $M_{1}$ represent the classical messages 0 and 1, respectively. Then, Alice send the pulses to Bob.

Step 4: Second round of security checking.  Bob selects the $X$, $Y$, or $Z$ bases to decode the received pulse based on the initial pulse base information. Then Alice announces the position of the randomly encoding pulses and the randomly encoded information, and Bob estimates the error rate of the second-round of transmission $E2$ based on the measurement results of the corresponding position. If $E2$ is below the tolerate threshold, the communication continues. Otherwise, the parties have to discard the communication.

Step 5: Decoding.  Bob deduces Alice's encoded information from the measurement results of the remaining photons and his own initial states.

\section{ The theoretical secrecy message capacity}
According to Wyner's wiretap channel theory \cite{Wyner}, the secrecy message capacity can be defined as
\begin{eqnarray}
	C_{s}&=&\max_{\{p_{0}\}} \{I(A:B)-I(A:E)\},
\end{eqnarray}
where $p_{0}$ is the probability that Alice encodes message $0$ with $M_{0}$ operation. $I(A:B)$ is the amount of information entropy that Bob can obtain through measurements from the photons encoded by Alice. It can be expressed as the product of the information entropy transmittable by a single pulse and the overall gain.  Since Alice performs two operations, the theoretical information entropy transmittable by a unit pulse is 1 bit. Subtracting the information loss due to errors  yields the correct information entropy transmittable by a unit pulse. $I(A:B)$ can be expressed as
\begin{eqnarray}
	I(A:B)=\langle    Q^{BAB}\rangle_{S_{K}^{s}} [1-h(\langle E^{BAB} \rangle_{S_{K}^{s}})], 
\end{eqnarray}
where $\langle    Q^{BAB}\rangle_{S_{K}^{s}}$ is  the overall gain of $K$ basis in $signal$ state interval. $\langle E^{BAB} \rangle_{S_{K}^{s}}$ is the error rate of $K$ basis in $signal$ state interval. 
$h(x)$ is the binary Shannon entropy function, and  $h(x)=-x\log_{2}(x)-(1-x)\log_{2}(1-x)$.

$I(A:E)$ is the amount of information entropy that Eve can obtain through attacks against the photons encoded by Alice. It can be expressed as the product of Eve's total gain and the information entropy Eve obtains from a single pulse. Considering photon number splitting (PNS) attack and collective attack, $I(A:E)$ can be decomposed into the product of the $n$-photon gain and the amount of information stolen from the $n$-photon state, and can be expressed as
\begin{eqnarray}
	I(A:E)=\sum_{n=0}^{\infty } Q_{n,s}^{BAE} H_{n}. 
\end{eqnarray}
$Q_{n,s}^{BAE}$  represent the overall gain of $n$-photon signal state attack by Eve.  According to \cite{decoy3}, it will be bounded by the overall gain of Alice during the first round transmission and we have
\begin{eqnarray}
	Q_{s,n }^{BAE}
	\le [\langle    Q^{BA}_{n}\rangle_{S_{K}^{s}} -\langle P_{I}(n)\rangle_{S_{K}^{s}} Y_{0}^{A}]max\{1,\frac{\gamma ^{E}}{\gamma^{A}}\},
\end{eqnarray}
where $\langle Q^{BA}_{n}\rangle_{S_{K}^{s}}$ is the overall gain of the $n$-photon $signal$ state during the first-round transmission. $\langle P_{I}(n)\rangle_{S_{K}^{s}}$ is the probability of the $n$-photon $signal$ state. $Y_{0}^{A}$ is the zero-photon yield  at Alice in the first transmission round.
 $\gamma ^{E}$  is the overall transmission efficiency of Eve after Alice encodes her receiving photons, including the channel  efficiency in the first-round transmission $t^{BA}$ and the optical device efficiency $\eta_{opt} ^{E}$ introduced  by Eve's attack. Then, after memory   and encoding   by Alice, Eve can steal the information of the photons emitted by Alice. $\eta_{m} ^{A}$ is the memory efficiency of Alice, and  $\eta_{opt} ^{A}$ is the optical device efficiency caused by Alice's encoding  operation. 
$\gamma^{A}$ is the overall transmission efficiency for photons received and then measured by Alice, including the  channel efficiency in the first-round transmission $t^{BA}$ and the detection efficiency $\eta_{D} ^{A}$ of Alice. 

$H_{n}$ represents the amount of information stolen from the $n$-photon state by Eve. According to Eq. (\ref{n}), the density matrix of a $n$-photon state can be expressed as
\begin{eqnarray}
	&&	|n\rangle_{a^{\dagger}(\theta,\phi)}\langle n| \\ \nonumber
	=&& \frac{1}{n! }\sum_{m_{1}=0}^{n} \sum_{m_{2}=0}^{n}C_{n}^{m_{1}} C_{n}^{m_{2}}  
	(\cos\frac{\theta}{2})^{m_{1}+m_{2}} (\sin\frac{\theta}{2})^{2n-m_{1}-m_{2}}\\ \nonumber 
	&&
	\sqrt{m_{1}!(n-m_{1})!}\sqrt{m_{2}!(n-m_{2})!} e^{-i\phi(m_{1}-m_{2})}\\ \nonumber 
	&& | m_{1} \rangle_{H}| n-m_{1} \rangle_{V}\langle m_{2}|_{H}\langle n-m_{2} |_{V}.
\end{eqnarray}
Due to the random distribution of the output state parameters $\theta$ and $\phi$ over the interval, integrating over the post-selected interval allows us to obtain the final output state as
\begin{eqnarray}
	\rho_{n}&&=\iiint _{\sum _{K,i}S_{K}^{i}}| n\rangle_{a^{\dagger}(\theta,\phi)}\langle n|
	f(I,\theta,\phi )dId\theta d\phi \\ \nonumber
	&&=\iiint _{\sum _{K,i}S_{K}^{i}}| n\rangle_{a^{\dagger}(\theta,\phi)}\langle n|
	f(I,\theta )\frac{1}{2 \bigtriangleup \phi}dId\theta d\phi \\ \nonumber
	&&=\rho_{n,Z}+\rho_{n,X}+\rho_{n,Y}.
\end{eqnarray}

Since the $0$-photon states cannot encode information, Eve cannot steal information from the $0$-photon states, and $H_{0}=0$.

The density matrix of the single-photon  states output by Bob is
\begin{eqnarray}\label{rou1}
	\rho_{1}&&=\rho_{1,X}+\rho_{1,Y}+\rho_{1,Z}=\frac{1}{2}(| H \rangle\langle H|+| V \rangle\langle V|),
\end{eqnarray}
which is the same as that in QSDC using a coherent light source. According to \cite{decoy3}, when considering collective attacks and PNS attacks, Eve is able to steal $h(e_{X}+e_{Z})$  information from single-photon errors, where $e_{X}(e_{Z})$ is the phase (bit) error rate of single photons. Suppose the bit and phase error rates of single photons are equal, we have $e_{X}=e_{Z}=e_{1}$ and $H_{1}=h(2e_{1})$.

The density matrices of the two-photon  states output by Bob is
\begin{eqnarray}\label{rou2}
	\rho_{2}&&=\rho_{2,X}+\rho_{2,Y}+\rho_{2,Z}\\ \nonumber 
	&&=(\int_{0 }^{I_{s}}dI  
	\int_{0 }^{\bigtriangleup }d\theta f(I,\theta )(\frac{1+\cos ^{2}\theta }{2} 
	(| HH \rangle\langle HH|\\ \nonumber 
	&&+| VV \rangle\langle VV|)+\sin ^{2}\theta | HV \rangle\langle HV|)\\ \nonumber 
	&&
	+2\int_{0 }^{I_{s}}dI  
	\int_{0 }^{\bigtriangleup }d\theta f(I,\frac{\pi}{2}-\theta )(\frac{1+\sin ^{2}\theta }{2} 
	(| HH \rangle\langle HH|\\ \nonumber 
	&&+| VV \rangle\langle VV|)+\cos ^{2}\theta | HV \rangle\langle HV|))/\\ \nonumber
	&&( \iint_{\sum S_{K}^{i} }  f(I,\theta ) dI d\theta ),
\end{eqnarray}
where $\bigtriangleup \theta_{X}=\bigtriangleup \theta_{Y}=\bigtriangleup \theta_{Z}=\bigtriangleup$.
It is evident that the two-photon states emitted by Bob are highly correlated with both   $I$  and  $\theta$, making it extremely difficult to estimate the amount of information Eve has stolen. Therefore, we consider the worst-case scenario, assuming that all information in the two-photon states is stolen by Eve, that is, $H_{2}=1$.  The derivation of Eqs. (\ref{rou1}) and (\ref{rou2}) are in the Appendix \ref{rou}.

For pulses with three or more photons, due to unambiguous discrimination attack \cite{Unambiguous}, Eve can steal all the information in the pulse, that is, $H_{n\ge 3}=1$.

Therefore, the information entropy that Eve can obtain through attacks can be defined as
\begin{eqnarray}
	I(A:E)= Q_{n=1,s}^{BAE} h(2e_{1}) +Q_{\ge2,s}^{BAE}\cdot1. 
\end{eqnarray}

Thus, the secrecy message capacity under collective attacks and PNS attacks can be defined as
\begin{eqnarray}\label{C}
	C_{K,s}=&& \langle    Q^{BAB}\rangle_{S_{K}^{s}} [1-h(\langle E^{BAB} \rangle_{S_{K}^{s}})]- \\ \nonumber
	&&	Q_{n=1,s}^{BAE} h(2e_{1})-Q_{\ge2,s}^{BAE}\cdot1.
\end{eqnarray}

\section{System model}
In this section, we establish a system model to analyze the performance of the passively-sourced QSDC protocol.

\begin{figure}[!htbp]
	\begin{center}
		\includegraphics[width=8cm,angle=0]{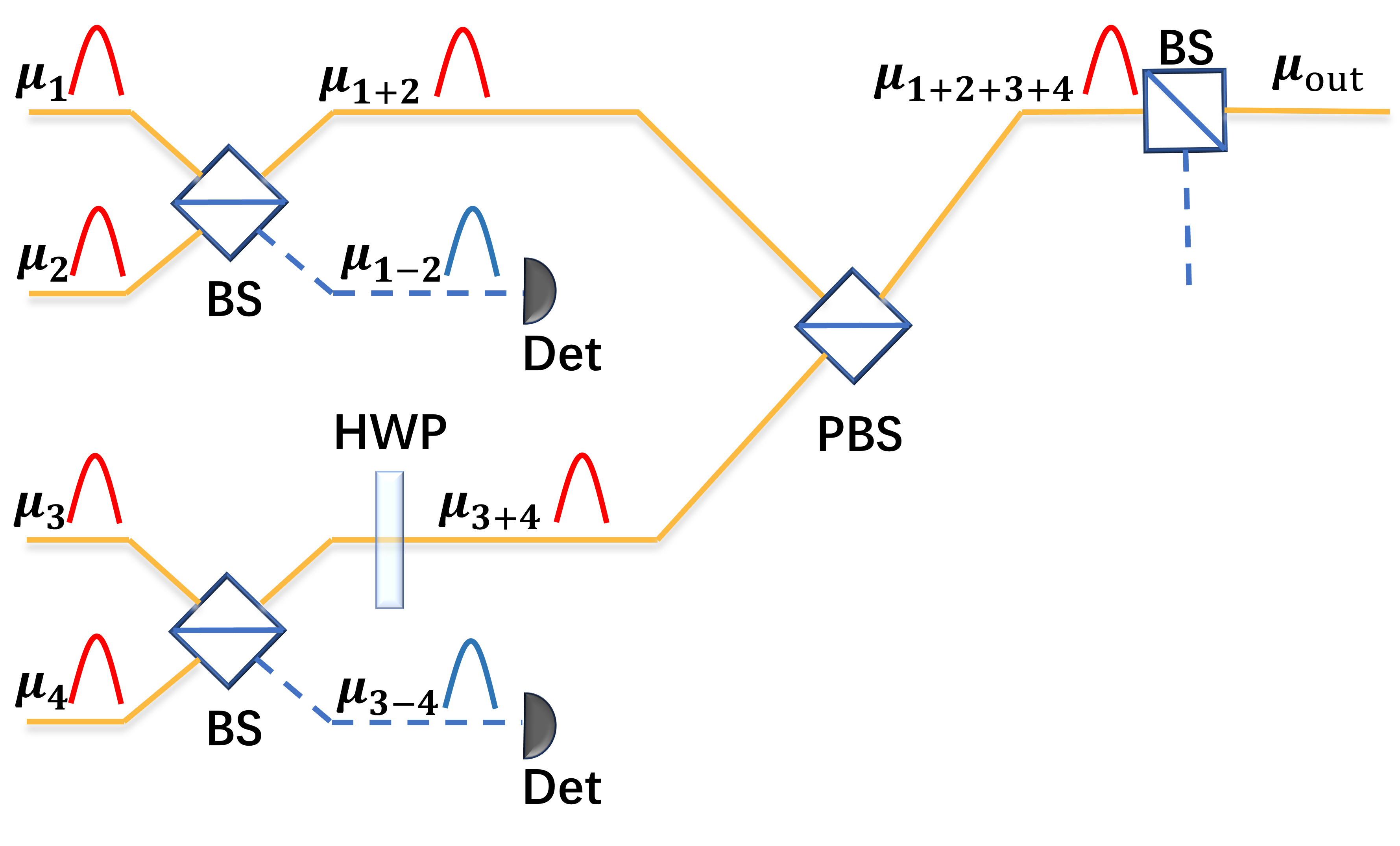}
		\caption{The structure of the symmetric fully passive source, which is similar to that in \cite{passive3}. BS: beam splitter; PBS:  polarization beam splitter; HWP:half-wave plate ; Det: detector. } \label{source}
	\end{center}
\end{figure}

\subsection{The fully passive source }\label{fpsource}
In the protocol, we use a fully passive source \cite{passive3}, whose structure is shown in the Fig. \ref{source}.
Among them, the initial four coherent pulses are as follows
\begin{eqnarray}
	| \mu_{1}   \rangle &=& | \sqrt{v}e^{i\alpha }    \rangle _{H},| \mu_{2}   \rangle = | \sqrt{v}e^{i\beta  }    \rangle _{H},\\  \nonumber
	| \mu_{3}   \rangle &=& | \sqrt{v}e^{i\gamma  }    \rangle _{H},| \mu_{4}   \rangle =| \sqrt{v}e^{i\delta   }    \rangle _{H},
\end{eqnarray}
where  $\alpha, \beta, \gamma,$ and $\delta $ are random phases of the four pulses, and the pulses have the same intensity $v$.
The subscript $H$  represents the quantum state of the photons in the pulse, corresponding to the horizontal polarization of the photons state $ | H   \rangle $. 

Firstly, the pulses $| \mu_{1}   \rangle$ and  $| \mu_{2}   \rangle$ are input into a beam splitter (BS), and the outputs $| \mu_{1+2}   \rangle$ and  $| \mu_{1-2}   \rangle$ can be derived as follows
\begin{eqnarray}
	&&| \mu_{1}   \rangle| \mu_{2}   \rangle =| \sqrt{v}e^{i\alpha }    \rangle _{H} | \sqrt{v}e^{i\beta  }  \rangle _{H}\\ \nonumber
	&&\overset{BS}{\rightarrow}| \sqrt{\frac{v}{2} }(e^{i\alpha} +e^{i\beta} )   \rangle _{H}
	| \sqrt{\frac{v}{2} }(e^{i\alpha} -e^{i\beta} )   \rangle _{H} \\  \nonumber
	&&=| \sqrt{2v }e^{i\frac{\alpha+\beta}{2} } \cos\frac{\alpha-\beta}{2}     \rangle _{H}
	| \sqrt{2v }e^{i\frac{\alpha+\beta+\pi }{2} } \sin\frac{\alpha-\beta}{2}     \rangle _{H} \\ \nonumber
	&&=| \mu_{1+2}   \rangle| \mu_{1-2}   \rangle.
\end{eqnarray}

Similarly,  $| \mu_{3+4}   \rangle$ and  $| \mu_{3-4}   \rangle$ can be written as
\begin{eqnarray}
	&&| \mu_{3+4}   \rangle=| \sqrt{2v }e^{i\frac{\gamma +\delta }{2} } \cos\frac{\gamma -\delta}{2}  \rangle _{H}, \\
	&&| \mu_{3-4}   \rangle=| \sqrt{2v }e^{i\frac{\gamma +\delta+\pi }{2} } \sin\frac{\gamma -\delta}{2}     \rangle _{H}.
\end{eqnarray}

The pulses $| \mu_{1-2}   \rangle$ and $| \mu_{3-4}   \rangle$ are used to obtain the source information for post-selection.
The pulse  $| \mu_{3+4}   \rangle$ first passes through a half-wave plate (HWP) to change the polarization state from horizontal ($|H\rangle$) to vertical ($|V\rangle$). Then, the changed pulse $| \mu_{3+4}^{'}   \rangle$ and  $| \mu_{1+2}   \rangle$ are respectively input into the two ports of a polarization beam splitter (PBS) for beam combining.
 After this, the output can be derived as follows
\begin{eqnarray}\label{s1}
	&&| \mu_{1+2}   \rangle| \mu_{3+4}^{'}   \rangle \overset{PBS}{\rightarrow}| \mu_{1+2+3+4}   \rangle\\\nonumber
	=&&| \sqrt{2v }e^{i\frac{\alpha+\beta }{2} } \cos\frac{\alpha-\beta}{2}     \rangle _{H}
	| \sqrt{2v }e^{i\frac{\gamma +\delta }{2} } \cos\frac{\gamma -\delta}{2}     \rangle _{V}\\\nonumber
	=&&\exp(-v\cos^{2}\frac{\alpha-\beta}{2})\exp(\sqrt{2v }e^{i\frac{\alpha+\beta }{2} } \cos\frac{\alpha-\beta}{2} a_{H}^{\dagger } )\\ \nonumber
	&&\exp(-v\cos^{2}\frac{\gamma -\delta}{2})\exp(\sqrt{2v }e^{i\frac{\gamma +\delta}{2} } \cos\frac{\gamma -\delta}{2} a_{V}^{\dagger } )| vac   \rangle \\\nonumber
	=&&\exp(-v(\cos^{2}\frac{\alpha-\beta}{2}+\cos^{2}\frac{\gamma -\delta}{2}))\\ \nonumber
	&&\exp(\sqrt{2v} e^{i\frac{\alpha+\beta }{2}(\cos\frac{\alpha-\beta}{2} a_{H}^{\dagger }+e^{i(\frac{\gamma +\delta}{2}-\frac{\alpha+\beta}{2})}\cos\frac{\gamma -\delta}{2} a_{V}^{\dagger })})
| vac   \rangle . \nonumber
\end{eqnarray}
Here, we make some substitutions, as follows
\begin{eqnarray}\label{itp}
	I_{l}&&=2v(\cos^{2}\frac{\alpha-\beta}{2}+\cos^{2}\frac{\gamma -\delta}{2}),\\\nonumber
	\cos\frac{\theta }{2} &&=\cos\frac{\alpha-\beta}{2}/ \sqrt{\cos^{2}\frac{\alpha-\beta}{2}+\cos^{2}\frac{\gamma -\delta}{2}},   \\\nonumber
	\sin\frac{\theta }{2} &&=\cos\frac{\gamma -\delta}{2}/ \sqrt{\cos^{2}\frac{\alpha-\beta}{2}+\cos^{2}\frac{\gamma -\delta}{2}},   \\\nonumber
	\psi &&=\frac{\alpha+\beta }{2},\\\nonumber
	\phi &&=\frac{\gamma +\delta}{2}-\frac{\alpha+\beta}{2},\\\nonumber
	a^{\dagger } (\theta , \phi )&&=\cos\frac{\theta }{2}a_{H}^{\dagger }+e^{i\phi}\sin\frac{\theta }{2}a_{V}^{\dagger }.
\end{eqnarray}
And we have
\begin{eqnarray}
	| \mu_{1+2+3+4}   \rangle&&=e^{-\frac{I_{l}}{2} }e^{\sqrt{I_{l}}e^{i\psi }a^{\dagger } (\theta , \phi ) }| vac   \rangle=| \sqrt{I_{l}}e^{i\psi }  \rangle_{a^{\dagger } (\theta , \phi )}, \quad\quad
\end{eqnarray}

Finally, the coherent pulse $| \mu_{1-2+3-4}   \rangle$ is attenuated to $| \mu_{out}   \rangle $ by BS, and
\begin{eqnarray}\label{muout}
	| \mu_{out}   \rangle&&=e^{-\frac{I}{2} }e^{\sqrt{I}e^{i\psi }a^{\dagger } (\theta , \phi ) }| vac   \rangle=| \sqrt{I}e^{i\psi }  \rangle_{a^{\dagger } (\theta , \phi )},
\end{eqnarray}
 where $ I=tI_{l}$ and $t$ is the transmission coefficient of BS. Since the output states are randomly distributed with respect to the azimuthal angle $\theta$ and $\phi$ on the Bloch sphere, we need to define post-selection intervals to filter out the target states for transmission.
We can define the polarization state  intervals and  decoy state intensity intervals  as
 \begin{eqnarray}
	S_{Z,z}^{i}=\{&&\phi \in( 0, 2\pi ) ,\\\nonumber
	&&\theta \in ( \theta_{z}-\bigtriangleup \theta_{Z} , \theta_{z}+\bigtriangleup \theta_{Z} ) \cap (0,\pi ),\\\nonumber
	&&i\in \{d1,d2,s\}\}.\\
	S_{X,x}^{i}=\{&&\phi \in( \phi _{x} -\bigtriangleup \phi_{X} , \phi _{x} +\bigtriangleup \phi_{X} ) ,\\\nonumber
	&&\theta \in ( \frac{\pi}{2}  -\bigtriangleup \theta_{X}   , \frac{\pi}{2} +\bigtriangleup \theta_{X} ) ,\\\nonumber
	&&i\in \{d1,d2,s\}\}.\\
	S_{Y,y}^{i}=\{&&\phi \in( \phi _{y} -\bigtriangleup \phi_{Y} , \phi _{y} +\bigtriangleup \phi_{Y} ) ,\\\nonumber
	&&\theta \in ( \frac{\pi}{2}  -\bigtriangleup \theta_{Y}   , \frac{\pi}{2} +\bigtriangleup \theta_{Y} ) ,\\\nonumber
	&&i\in \{d1,d2,s\}\}.
\end{eqnarray}
where the superscript $i$ represents the decoy state intensity intervals, and  $d1,d2,s$ correspond  three decoy state intensity intervals where $s \in (I_{d},I_{s}],d2\in (I_{vac},I_{d}] ,d1\in [0,I_{vac}]  $, $I_{vac}$, $I_{d}$, and $I_{s} $ are the boundaries of three intervals.
The subscript $X$, $Y$, and $Z$ denote different bases, while $x$, $y$, and $z$  correspond to different states.
Here, $x\in \{D,A\},y\in \{R,L\},z\in \{H,V\}$, which respectively correspond to the diagonal ($D$), anti-diagonal ($A$), right-circularly ($R$), left-circularly ($L$), horizontal ($H$), and vertical ($V$) polarization states of photons.  $S_{Z}^{i}= S_{Z,H}^{i} \cup S_{Z,V}^{i}$, $S_{X}^{i}= S_{X,D}^{i} \cup S_{X,A}^{i}$, and $S_{Y}^{i}= S_{Y,R}^{i} \cup S_{Y,L}^{i}$.
When $\theta_{z}=0, \pi$, $z$ corresponds to states  $H$, and $V$, respectively. 
When $\phi_{x}=0, \pi$,  $x$ corresponds to states  $D$, and $A$, respectively. 
When $\phi_{y}=\frac{\pi}{2}, \frac{3\pi}{2}$,  $y$ corresponds to states  $R$, and $L$, respectively. 
 $\bigtriangleup \theta_{X}$,$\bigtriangleup \theta_{Y}$, $\bigtriangleup \theta_{Z}$, $\bigtriangleup \phi_{Y}$, and $\bigtriangleup \phi_{Z}$ characterize the size of the interval.
 


\subsection{Photon transmission efficiency}
In the whole system, photons are mainly lost by quantum channels, optical devices, and photon detector. Here, we consider the first two factors, which the photon detector will analyze in the next subsection.
 QSDC requires two rounds of transmission. Here, we denote the first round of Bob to Alice as $BA$, and the second round of Alice to Bob as $BAB$. Then, the transmission efficiency of the system can be written as
\begin{eqnarray}
	\eta ^{chan}&= &t^{chan}\eta_{opt} ^{chan},
\end{eqnarray}
where $chan\in\{BA,BAB\}$, $t^{chan}$ is the channel transmission efficiency,  $\eta_{opt} ^{chan}$  is the intrinsic optical efficiency of the devices.

Considering the most common fiber loss channels, $t^{chan}$ can be further written as
\begin{eqnarray}
	t^{chan}=10^{-\frac{\alpha ^{chan}}{10} },
\end{eqnarray}
where $\alpha ^{chan}$ is quantum channel attenuation corresponding to the transmission round, and it is typically the product of the fiber loss coefficient ($\alpha_{f}$) and the photon transmission distance ($L$).

\subsection{overall gain $Q$ and error rate $E$} \label{QEtit}
Next, we need to estimate overall gain $Q$ and error rate $E$. 
According to Eq. (\ref{muout}), before integrating over the interval, the output state generated by the fully passive source is a phase-randomized coherent light pulse. We can calculate the gains of the quantum states corresponding to parameters $I$, $\theta$, and $\phi$ through the system model, and then integrate over the interval to obtain the total gain of the quantum states within the interval.
First, we consider the fully passive source  emits $n$ photons with a probability of  $P_{I}(n)$ where $P_{I } (n)  = e^{-I }\frac{I ^{n}}{n!}$ , and the photons are in the quantum state $a^{\dagger } (\theta , \phi )$.
Here we define   the probability that $n$ photons arrive at the detector as $P_{n}$  and $P_{n}$ is shown as follows
\begin{eqnarray}
	P_{n } & &=  \sum_{N  =  n}^{\infty } P_{I}(N)S(N,n)\\ \nonumber
	&&=\sum_{N  =  n}^{\infty } e^{-I }\frac{I ^{N}}{N!}C_{N}^{n} (1-\eta ^{chan})^{N-n}(\eta ^{chan})^{n}\\\nonumber
	&&=\sum_{N  =  n}^{\infty } e^{-I }\frac{I ^{N}}{N!}\frac{N!}{(N-n)!n!}  (1-\eta ^{chan})^{N-n}(\eta ^{chan})^{n}\\\nonumber
	&&\overset{t=N-n}{\rightarrow} \sum_{t  =  0}^{\infty }I^{n}e^{-I }\frac{I ^{t}}{t!n!}(1-\eta ^{chan})^{t}(\eta ^{chan})^{n}\\\nonumber
	&&=\frac{(I\eta ^{chan})^{n}}{n!} e^{-I\eta ^{chan} },
\end{eqnarray}
where $S(N,n)$ represents that the source emits $N$ photons and $n$ photons arrive at the detector.

When the $n$ photons arrive at the detector, we have
\begin{eqnarray}\label{nn}
	| n \rangle _{a^{\dagger } (\theta , \phi )}=&&\frac{1}{\sqrt{n!} }
	(a^{\dagger } (\theta , \phi ))^{n} | vac \rangle \label{n} \\ \nonumber
	=&&\frac{1}{\sqrt{n!} }(\cos\frac{\theta}{2} a_{H}^{\dagger }+
	e^{i\phi  }\sin\frac{\theta}{2}a_{V}^{\dagger })^{n}| vac \rangle\\ \nonumber
	=&&\frac{1}{\sqrt{n!} }(f(k) a_{k}^{\dagger }+
	f(l)a_{l}^{\dagger })^{n}| vac \rangle\\ \nonumber
	=&& \frac{1}{\sqrt{n!} }\sum_{m=0}^{n} C_{n}^{m}(f(k)a_{k}^{\dagger })^{m}(f(l)a_{l}^{\dagger })^{n-m}| vac \rangle\\ \nonumber
	=&&\frac{1}{\sqrt{n!} }\sum_{m=0}^{n} C_{n}^{m} f(k)^{m}
	f(l)^{n-m} \\ \nonumber
	&&\sqrt{m!(n-m)!}  | m \rangle_{k}| n-m \rangle_{l},\\
	(k,l)&&\in  \{ (H,V),(D,A),(R,L)  \} \label{x}.
\end{eqnarray}
Notice that
\begin{eqnarray}
	a^{\dagger } (\theta , \phi )=&&\cos\frac{\theta}{2} a_{H}^{\dagger }+
	e^{i\phi  }\sin\frac{\theta}{2}a_{V}^{\dagger }   \\ \nonumber
	=&&f(H)a_{H}^{\dagger }+f(V)a_{V}^{\dagger }\\ \nonumber
	=&&\frac{1}{\sqrt{2} }(\cos\frac{\theta}{2}+e^{i\phi  }\sin\frac{\theta}{2})a_{D}^{\dagger }+\\ \nonumber
	&&\frac{1}{\sqrt{2} }(\cos\frac{\theta}{2}-e^{i\phi  }\sin\frac{\theta}{2})a_{A}^{\dagger }\\ \nonumber
	=&&f(D)a_{D}^{\dagger }+f(A)a_{A}^{\dagger }\\ \nonumber
	=&&\frac{1}{\sqrt{2} }(\cos\frac{\theta}{2}-ie^{i\phi }\sin\frac{\theta}{2})a_{R}^{\dagger }+\\ \nonumber
	&&\frac{1}{\sqrt{2} }(\cos\frac{\theta}{2}+ie^{i\phi  }\sin\frac{\theta}{2})a_{L}^{\dagger }\\ \nonumber
	=&&f(R)a_{R}^{\dagger }+f(L)a_{L}^{\dagger },\\ \nonumber
\end{eqnarray}
and
\begin{eqnarray}
	&|f(H)| ^{2}=\frac{1+\cos\theta }{2} ,&|f(V)| ^{2}=\frac{1-\cos\theta }{2} ,\\ \nonumber
	&|f(D)| ^{2}=\frac{1+\sin\theta \cos\phi}{2} ,&|f(A)| ^{2}=\frac{1-\sin\theta \cos\phi}{2}, \\ \nonumber
	&|f(R)| ^{2}=\frac{1+\sin\theta \sin\phi}{2} ,&|f(L)| ^{2}=\frac{1-\sin\theta \sin\phi}{2}. \\ \nonumber
\end{eqnarray}
Then, According to Eq. (\ref{nn}), the probability of the state $| m \rangle_{k}| n-m \rangle_{l}$ is
\begin{eqnarray}
	P_{kl}(m,n-m)=C_{n}^{m} |f(k)|  ^{2m}|f(l)| ^{2(n-m)}.
\end{eqnarray}

In this way, the probability that $n$ photons arrive at detector and trigger response to state $k$  can be expressed as
\begin{eqnarray} \label{PH}
	P_{k}^{n} &&=\sum_{m=0 }^{n} P_{kl}(m,n-m)[1-(1-\eta_{D})^{m}(1-Pd)]\\ \nonumber
	&&\quad (1-\eta_{D})^{n-m}(1-Pd)\\\nonumber
	&&=\sum_{m=0 }^{n}C_{n}^{m} |f(k)|  ^{2m}|f(l)| ^{2(n-m)}\\\nonumber
	&&\quad	[1-(1-\eta_{D})^{m}(1-Pd)](1-\eta_{D})^{n-m}(1-Pd)\\\nonumber
	&&=(1-Pd)(|f(k)|^{2}+|f(l)|^{2}(1-\eta_{D}))^{n}-\\\nonumber
	&&\quad(1-Pd)^{2}(|f(k)|^{2}(1-\eta_{D})+|f(l)|^{2}(1-\eta_{D}))^{n}\\\nonumber
	&&=(1-Pd)(1-|f(l)|^{2}\eta_{D})^{n}-(1-Pd)^{2}(1-\eta_{D})^{n},
\end{eqnarray}
where $\eta_{D}$ is the detection efficiency, and $Pd$ is the dark count probability.

Similarly, the probability that $n$ photons arrive at detector and trigger response to state $l$  can be expressed as
\begin{eqnarray}\label{PV}
	P_{l}^{n} &&=(1-Pd)(1-|f(k)|^{2}\eta_{D})^{n}-(1-Pd)^{2}(1-\eta_{D})^{n}. \quad\quad
\end{eqnarray}

Then, the gain of the  state $k$ and $l$ can be calculated as
\begin{eqnarray}
	&&Q_{k}^{chan}=\sum_{n=0 }^{\infty}P_{n }	P_{k}^{n}\\ \nonumber
	&&=\sum_{n=0 }^{\infty}\frac{(I\eta ^{chan})^{n}}{n!} e^{-I\eta ^{chan} }
	[(1-Pd)(1-|f(l)|^{2}\eta_{D})^{n}- \\ \nonumber
&&\quad	(1-Pd)^{2}(1-\eta_{D})^{n}]\\\nonumber
	&&=(1-Pd)e^{-I\eta ^{chan}|f(l)|^{2}\eta_{D}}-(1-Pd)^{2}e^{-I\eta ^{chan} \eta_{D}},\\\nonumber
	&&Q_{l}^{chan}=\sum_{n=0 }^{\infty}P_{n }P_{l}^{n}\\\nonumber
	&&=(1-Pd)e^{-I\eta ^{chan}|f(k)|^{2}\eta_{D}}-(1-Pd)^{2}e^{-I\eta ^{chan} \eta_{D}}.
\end{eqnarray}

Finally, the overall gain of the state $a^{\dagger } (\theta , \phi )$ can be expressed as
\begin{eqnarray}
	Q^{chan} & = & Q_{k}^{chan}+Q_{l}^{chan},\\
\end{eqnarray}
and for state $k$, the error rate can be expressed as
\begin{eqnarray}
E^{chan}_{k} & = &\frac{e_{d}Q_{k}^{chan}+(1-e_{d})Q_{l}^{chan}}{Q^{chan}},
\end{eqnarray}
where $e_{d}$ is the optical intrinsic error rate.

 Due to the randomness of the source, the quantum state of its output  will fluctuate in a range. So we need to integrate over the corresponding state interval and the corresponding intensity interval. Thus, the overall gain Q of the interval $S_{K,k}^{i}$ can be calculated as
\begin{eqnarray}
	\langle Q^{chan}\rangle_{S_{K,k}^{i}} =\frac{1}{	\langle P\rangle_{S_{K,k}^{i}}}\iiint_{S_{K,k}^{i}}^{}&& (Q_{k}^{chan}+Q_{l}^{chan})\nonumber\\
	&&f(I,\theta ,\phi )dI d\theta d\phi .
\end{eqnarray}

We can also calculate the qubit error rate E. Taking state k for example, we have
\begin{eqnarray}
	\langle E^{chan}\rangle_{S_{K,k}^{i}} =\frac{1}{	\langle P\rangle_{S_{K,k}^{i}}} \iiint_{S_{K,k}^{i}}^{}&& \frac{e_{d}Q_{k}^{chan}+(1-e_{d})Q_{l}^{chan}}{Q_{k}^{chan}+Q_{l}^{chan}}  \nonumber \\
	&&f(I,\theta ,\phi )dI d\theta d\phi .
\end{eqnarray}
where $K$ is the basis corresponding to state $k$ and $l$, and $K \in \{X,Y,Z\}$.
 $f(I, \theta ,\phi)$  is a probability density function \cite{passive5}.  $\langle P\rangle_{S_{K,k}^{i}}$ is the probability of selecting the interval, which is used for normalization, and
\begin{eqnarray}
	\langle P \rangle _{S_{K,k}^{i}}&&=\iiint_{S_{K,k}^{i}} f(I,\theta ,\phi )dId\theta d\phi , \label{pn}\\
		f(I,\theta ,\phi )&&=f(I,\theta  )f(\phi),\\
		f(\phi)&&=\frac{1}{2\Delta\phi   } ,\\
		f(I,\theta  )&&=\frac{1}{vt\pi ^{2}\sqrt{1-\frac{I}{2vt}\cos^{2}\frac{\theta}{2}  } \sqrt{1-\frac{I}{2vt}\sin^{2}\frac{\theta}{2}  }} .
\end{eqnarray}
More details in Appendix \ref{fun}.
Considering the symmetry, $\langle    Q^{BAB}\rangle_{S_{K}^{s}}=\langle    Q^{BAB}\rangle_{S_{K,k}^{s}}$ and $\langle E^{BAB} \rangle_{S_{K}^{s}}=\langle E^{BAB} \rangle_{S_{K,k}^{s}}$.

\section{Decoy state method}
Considering PNS attacks, Eve can steal information from the single-photon error rate. We need to estimate the single-photon yield $Y_{1}$ and error rate $e_{1}$ to bound Eve's eavesdropping.
According to Refs. \cite{passive5} and \cite{passive6},   we need to  solve the following linear programming problems
\begin{eqnarray}
	&&\min  \langle Y_{1}  \rangle _{S_{K}^{s}}, \label{y}\\ \nonumber
	s.t.&&\langle Q^{BA}\rangle_{S_{K}^{i}} \ge \sum_{n=0}^{n_{cut}} \langle P_{I}(n)\rangle_{S_{K}^{i}} \langle Y_{n} \rangle_{S_{K}^{i}},\\  \nonumber
	&&\langle Q^{BA}\rangle_{S_{K}^{i}} \le \sum_{n=0}^{n_{cut}} \langle P_{I}(n)\rangle_{S_{K}^{i}} \langle Y_{n}  \rangle_{S_{K}^{i}}\\  \nonumber
&&\quad\quad\quad\quad\quad\quad\quad	+  1-\sum_{n=0}^{n_{cut}} \langle P(I,n)\rangle_{S_{K}^{i}},\\  \nonumber
	&&|   \langle Y_{n}  \rangle _{S_{K}^{i}}-\langle Y_{n}  \rangle _{S_{K}^{j}}|
	\le D(\rho_{S_{K}^{i}}^{n}, \rho_{S_{K}^{j}}^{n}), n=2,...,n_{cut},\\  \nonumber
	&&\langle Y_{n}  \rangle _{S_{K}^{i}}=\langle Y_{K}^{n}  \rangle _{S_{K}^{i}}, n=0,1,\\  \nonumber
	&&0\le \langle Y_{n}   \rangle_{S_{K}^{i}} \le 1 , n=0,1,...,n_{cut},\\ \nonumber
	&&i,j\in \{d1,d2,s\},
\end{eqnarray}
\begin{eqnarray}
	&&\max   \langle e_{1}Y_{1}  \rangle _{S_{k}^{s}},\label{ey} \\  \nonumber
	s.t.&&\langle E^{BA}Q^{BA}  \rangle_{S_{k}^{j}} \ge \sum_{n=0}^{n_{cut}} \langle P_{I}(n)\rangle_{S_{k}^{j}} \langle e_{n}Y_{n}  \rangle_{S_{k}^{j}},\\  \nonumber
	&&\langle E^{BA}Q^{BA} \rangle_{S_{k}^{j}} \le \sum_{n=0}^{n_{cut}} \langle P_{I}(n)\rangle_{S_{k}^{j}} \langle e_{n}Y_{n}   \rangle_{S_{k}^{j}} \\  \nonumber
	&& \quad\quad\quad\quad\quad\quad\quad
	+ 1-\sum_{n=0}^{n_{cut}} \langle P_{I}(n)\rangle_{S_{k}^{j}},\\  \nonumber
	&&|   \langle e_{n}Y_{n}  \rangle _{S_{k}^{i}}-\langle e_{n}Y_{n}  \rangle _{S_{k}^{j}}|
	\le D(\rho_{S_{k}^{i}}^{n}, \rho_{S_{k}^{j}}^{n}),  n=1,...,n_{cut},\\  \nonumber
	&&\langle e_{0}Y_{0}  \rangle _{S_{k}^{i}}=\langle e_{0}Y_{0}  \rangle _{S_{k}^{j}},\\  \nonumber
	&&0\le \langle e_{n}Y_{n}  \rangle_{S_{k}^{j}} \le 1 , n=0,1,...,n_{cut},\\ \nonumber
	&&   i,j\in \{d1,d2,s\},
\end{eqnarray}
where $ D(\rho_{S_{K}^{i}}^{n}, \rho_{S_{K}^{j}}^{n})$ is the trace distance between density matrix $\rho_{S_{K}^{i}}^{n}$ and $ \rho_{S_{K}^{j}}^{n}$ , and is equal to $\sum |\lambda_{i}| $. $\lambda_{i}$ are the eigenvalues of density matrix  ($\rho_{S_{K}^{i}}^{n}- \rho_{S_{K}^{j}}^{n}$), and
\begin{eqnarray}
	\rho_{S_{K}^{i}}^{n}=\iiint_{S_{K}^{i}} |n\rangle\langle n|_{a^{\dagger}(\theta,\phi)}f(I,\theta,\phi)dId\theta d\phi.
\end{eqnarray}

\section{Numerical simulation}
In this section, we perform numerical simulations for the passively-sourced QSDC protocol. The parameters used are from the Pan's experiment \cite{decoy3}, shown as  Tab. \ref{para}.
\begin{table}[h]
	 \centering
	\vspace{-0.1cm}
	\setlength{\abovecaptionskip}{0.3cm}
	\setlength{\belowcaptionskip}{0.1cm}
	\setlength\tabcolsep{5pt} 
	\renewcommand\arraystretch{1.5}  
	\caption{Parameters used in the numerical simulation \cite{decoy3}.}
	\begin{tabular}{cccccc}  
		\hline
		\hline
	$\eta_{opt}^{BA}$ & $\eta_{opt}^{BAB}$ & Pd &$e_{d}^{A}$   & $e_{d}^{B}$   & $\eta_{D}$\\
		0.21    & 0.088    & $8 \times 10^{-8}$  &0.0131     & 0.0026    & 0.7  \\  \hline \hline
	\end{tabular}\label{para}
\end{table}

\begin{figure}[!htbp]\label{dis}
	\begin{center}
		\includegraphics[width=8cm,angle=0]{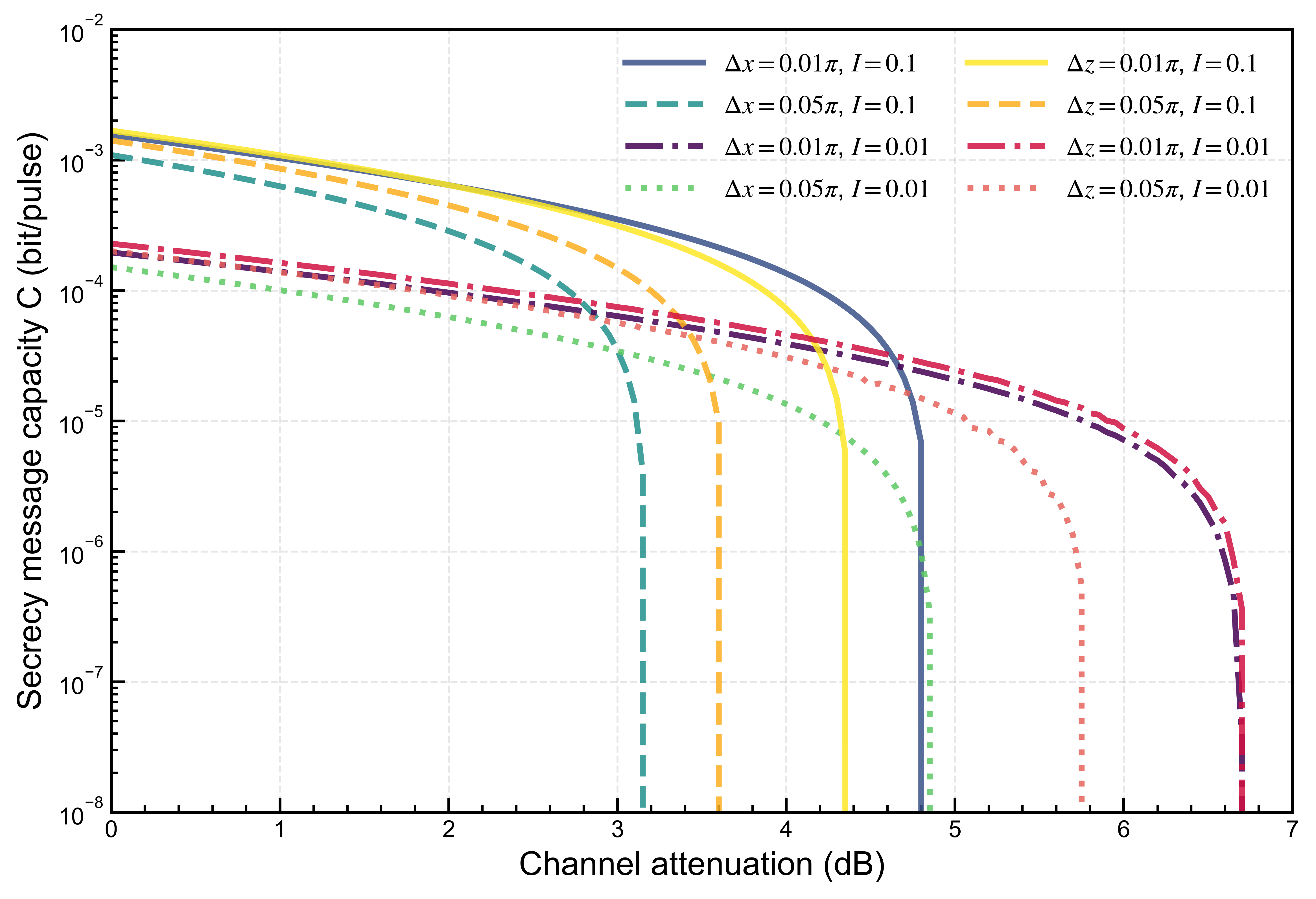}
		\caption{Comparison of the passively-sourced QSDC with different parameters.
		 Here, we set  $\bigtriangleup \theta_{Z}=\bigtriangleup z$ and
		 $\bigtriangleup \theta_{X}=\bigtriangleup \theta_{Y}=\bigtriangleup \phi_{X}=\bigtriangleup \phi_{Y}=\bigtriangleup x$.  
		 The four warm-color lines (yellow, orange, red, pink) represent the secrecy message capacity of the $Z$ basis.
		 The four cool-color lines (blue, cyan, purple, green) correspond to the secrecy message capacity  of the $X$ basis (and by symmetry, the $Y$ basis).
			 }\label{simulation1}
	\end{center}
\end{figure}

Fig. \ref{simulation1} illustrates the secrecy message capacity versus the channel attenuation under the collective attack as well as the PNS attack in the framework of decoy-state analysis. We compared the passively-sourced QSDC under different intervals. We set $\bigtriangleup \theta_{Z}=\bigtriangleup z$ and
$\bigtriangleup \theta_{X}=\bigtriangleup \theta_{Y}=\bigtriangleup \phi_{X}=\bigtriangleup \phi_{Y}=\bigtriangleup x$. The boundaries of three decoy-state intervals  are $I_{vac}=0.05I$, $I_{d}=0.1I$, and $I_{s}=I$.  
We have plotted two sets of curves corresponding to the $X$ basis and $Z$ basis, respectively. Due to symmetry, the results for the $Y$ basis are identical to those of the $X$ basis. We generated four curves by taking intensities of $0.1$ and $0.01$, and interval sizes of $0.01\pi$ and $0.05\pi$ for $X$ basis and $Z$ basis. Specifically,
the four warm-color lines (yellow, orange, red, pink) represent the secrecy message capacity of the $Z$ basis.
The four cool-color lines (blue, cyan, purple, green) correspond to the secrecy message capacity  of the $X$ basis.

From the Fig. \ref{simulation1}, it can be seen that the results of the $Z$ basis and $ X$ basis differ significantly in both secrecy message capacity and channel attenuation. This is caused by the shape of their post-selected intervals and the distinct probability distributions over different intervals. But in general,
the smaller the interval, the better the performance of secrecy message capacity and communication distance. This is because the fully passive source will generate imperfect states, and there are inherent deviations that introduce errors. As the interval increases, these deviations accumulate, leading to a rise in error rates and degrading the protocol's performance.
Furthermore, the variation law of the intensity  is consistent with that of actively modulated QSDC. Specifically,
in the low-loss regime, higher intensities yield larger secrecy message capacities.
In the high-loss regime, reducing the  pulse intensity leads to better performance.

In Fig. \ref{simulation1}, it appears that smaller intervals lead to better protocol performance. However, unlike traditional QSDC based on weak coherent pulses, in passively-sourced QSDC, higher pulse performance does not necessarily mean more information can be transmitted per unit time. This is because the fully passive sources rely on post-selection to generate output pulses, and smaller intervals result in lower post-selection success rates. Therefore, we define a secrecy message transmission rate to represent the protocol's performance, which is defined as $C_{s}=\sum_{K}\langle P \rangle _{S_{K}^{s}}C_{K}$. $\langle P \rangle _{S_{K}^{s}}$  is the probability of generating photons in the $K$ basis, and can be obtained by calculating the probabilities of the two states in the $K$ basis according to Eq. (\ref{pn}).  $C_{K}$ is the secrecy message capacity that each photon in the $K$ basis can transmit. 

\begin{figure}[!htbp]
	\begin{center}
		\includegraphics[width=9cm,angle=0]{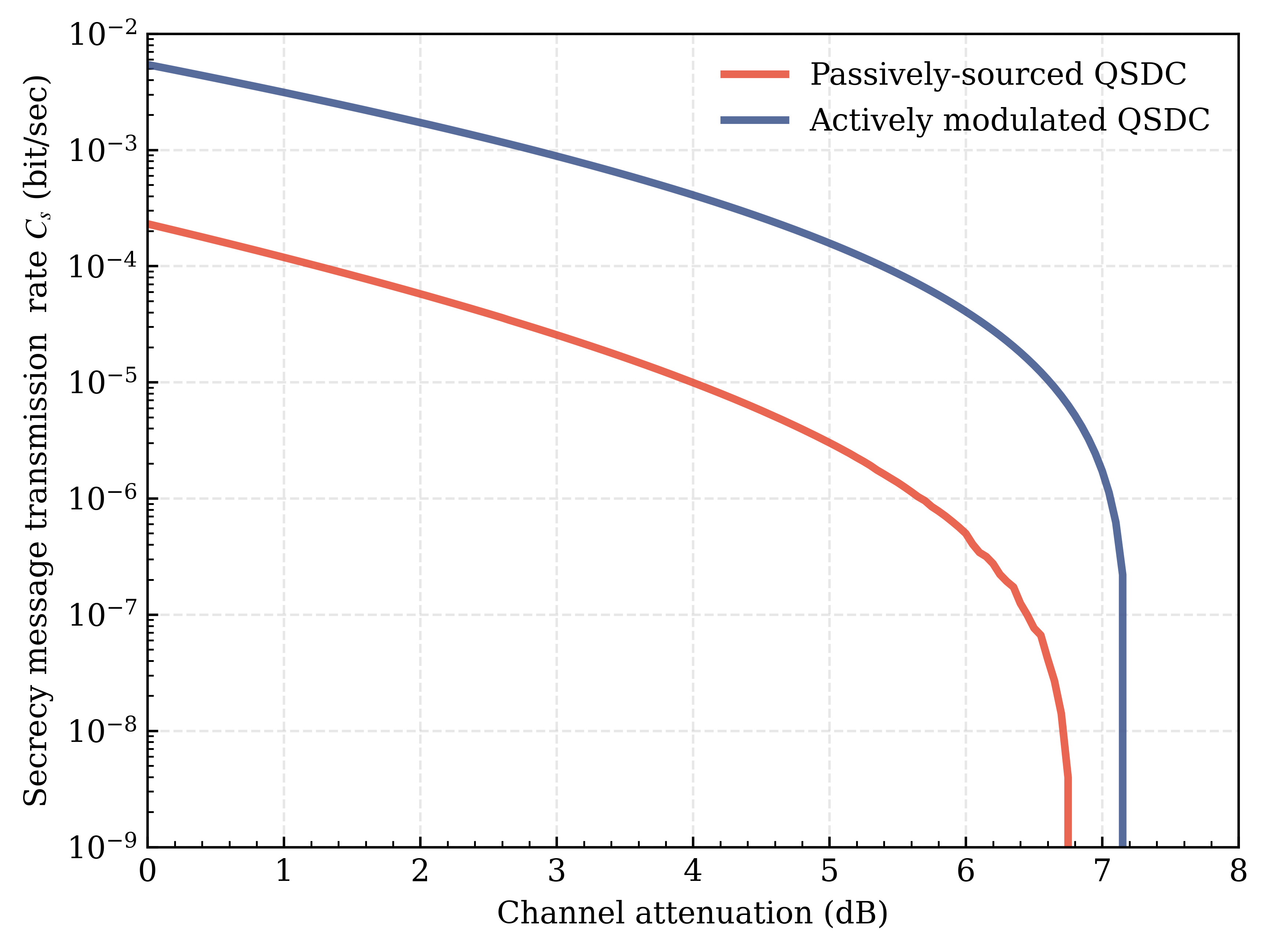}
		\caption{Comparison between passively-sourced QSDC and active-modulated QSDC at the optimal intensities and intervals. The red curve represents the passively-sourced QSDC, and the blue curve represents the actively modulated QSDC.}\label{s2}
	\end{center}
\end{figure}

\begin{figure}[!htbp]
	\begin{center}
		\includegraphics[width=9cm,angle=0]{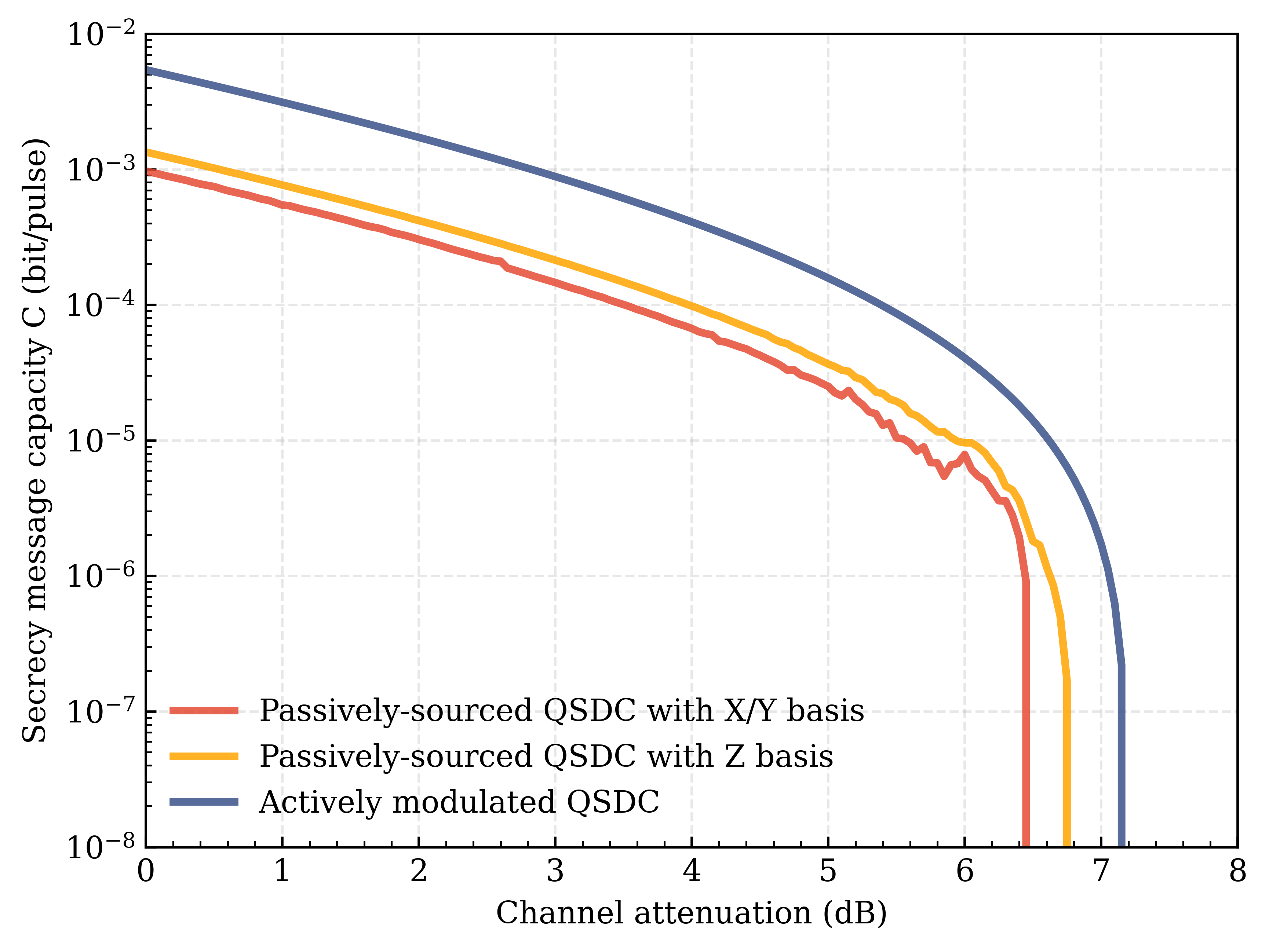}
		\caption{The secrecy message capacities of the protocol under $X$ basis and $Z$ basis when achieve the optimal secrecy message transmission rate in Fig. \ref{s2}. The red and orange curves correspond to the secrecy message capacities of the $X$ basis and $Z$ basis, respectively. The blue curve represents the secrecy message capacity of the actively modulated QSDC under optimal intensity.
		} \label{c1}
	\end{center}
\end{figure}

We performed parameter optimization on the protocol's secrecy message transmission rate, including intensity and interval sizes, to obtain the optimal performance for each channel attenuation and compared it with the optimal actively modulated QSDC. As shown in Fig. \ref{s2}, the red curve represents the passively-sourced QSDC, and the blue curve represents the actively modulated QSDC.
At a channel attenuation of 2, 4, 6 dB (corresponding to a communication distance of 5, 10, 15 km), the  secrecy message transmission rates of passively-sourced QSDC are $5.76 \times 10^{-5}$($I=0.0895, \bigtriangleup x=0.0490 \pi, \bigtriangleup z=0.0546  \pi$),  $9.92 \times10^{-6}$($I=0.0471, \bigtriangleup x=0.0367 \pi, \bigtriangleup z=0.0408 \pi$), and $4.99\times10^{-7} $($I=0.0168, \bigtriangleup x=0.0152 \pi, \bigtriangleup z=0.0214 \pi$) bit/sec. And its maximum communication distance is about $16.875 $ km. Compared with actively modulated QSDC, the secrecy message transmission rate of passively-sourced QSDC is approximately an order of magnitude lower. However, its communication distance can reach about $94.4\%$ of that of actively modulated QSDC.
This is partly due to the success rate of post-selection intervals affecting the protocol's efficiency, and partly due to errors introduced by imperfect sources, which degrade the protocol's pulse performance. In Fig. \ref{c1}, we also present the secrecy message capacities of the protocol under $X$ basis and $Z$ basis when achieving the optimal secrecy message transmission rate. The red and orange curves correspond to the secrecy message capacities of the $X$ basis and $Z$ basis, respectively. The blue curve represents the secrecy message capacity of the actively modulated QSDC under optimal intensity.





\section{Conclusion}

In conclusion, we propose a  QSDC protocol based on fully passive source. Weak coherent pulses with random intensity and random quantum state can be generated passively through the interference, combination and attenuation of four phase random coherent pulses.
This passive modulation can not only reduce experimental operations, but also resist side-channel attacks from third parties on the source modulators, thereby lowering the requirements for equipment.
 In terms of security analysis, we conducted decoy-state analysis considering PNS attacks and collective attacks of third-party Eve.
We analyzed the secrecy message capacity of the protocol under different parameters and defined a secure information transmission rate to quantify the information entropy that can be transmitted per unit time. We optimized the intensity and interval sizes of the fully passive source to obtain the protocol's maximum secrecy message transmission rate for each channel attenuation.
At a channel attenuation of 2, 4, 6 dB (corresponding to a communication distance of 5, 10, 15 km), the  secrecy message transmission rates of passively-sourced QSDC are $5.76 \times 10^{-5}$($I=0.0895, \bigtriangleup x=0.0490 \pi, \bigtriangleup z=0.0546  \pi$),  $9.92 \times10^{-6}$($I=0.0471, \bigtriangleup x=0.0367 \pi, \bigtriangleup z=0.0408 \pi$), and $4.99\times10^{-7} $($I=0.0168, \bigtriangleup x=0.0152 \pi, \bigtriangleup z=0.0214 \pi$) bit/sec.
 And its maximum communication distance is about $16.875 $ km, which is about $94.4\%$ of that of actively modulated QSDC.

\appendix
\section{probability density function} \label{fun}
Here we are talking about the probability density function, similar to \cite{passive5}.
First, we need an assumption that the initial four coherent pulse sources have phases that are uniform distribution over the domain.
According to Eq. (\ref{itp}), we can easily see that $\phi$ is also uniform distribution over the domain. And we have
\begin{eqnarray}
	f(I,\theta ,\phi )&&=f(I,\theta  )f(\phi),\\
	f(\phi)&&=\frac{1}{2\Delta\phi  }.
\end{eqnarray}

According to  Eq. (\ref{itp}),  let $\frac{\alpha-\beta}{2}$=$\theta_{1} $ and $\frac{\gamma -\delta}{2}$=$\theta_{2} $ , we have
\begin{eqnarray}
	&&I=2vt(\cos^{2}\theta _{1}+\cos^{2}\theta _{2}),\\\nonumber
	&&\theta=2\arctan\frac{\cos\theta _{2}}{\cos\theta _{1}}, \\\nonumber
	&&\cos\theta _{1}=\cos\frac{\theta }{2}\sqrt{\frac{I}{2vt} }, \\\nonumber
	&&\cos\theta _{2}=\sin\frac{\theta }{2}\sqrt{\frac{I}{2vt} }.
\end{eqnarray}
Obviously, $\theta_{1} $ and $\theta_{2} $ are  uniformly distributed. According to the transformation formula for random variables, we have
\begin{eqnarray}
	f(I,\theta  )=\frac{1}{ | J  | } f(\theta _{1},\theta _{2}),
\end{eqnarray}
where $| J  |$ is  Jacobian determinant, and
\begin{eqnarray}
	| J  | &&=\begin{vmatrix}
		\frac{ \partial I}{\partial\theta _{1}}  & \frac{ \partial I}{\partial\theta _{2}}\\
		\frac{ \partial \theta }{\partial\theta _{1}} & \frac{ \partial \theta }{\partial\theta _{2}}
	\end{vmatrix}=8vt\sin\theta _{1}\sin\theta _{2}\\
	&&=8vt\sqrt{1-\frac{I}{2vt}\cos^{2}\frac{\theta}{2}  } \sqrt{1-\frac{I}{2vt}\sin^{2}\frac{\theta}{2}  }.
\end{eqnarray}
$f(\theta _{1},\theta _{2})$ is a constant that depends on the domain.
The value of $\theta _{1}$ and $\theta _{2}$   affect the domain of $I$ and $\theta $ , but it is difficult to use $I$ and $\theta $ to backinfer the domain of $\theta _{1}$ and $\theta _{2}$. Therefore, we retrodict the probability density function according to its normalization conditions. Thus
\begin{eqnarray}
	&&\int_{I=0 }^{I=2vt} \int_{\theta=0}^{\theta=\pi } f(I,\theta  )dId\theta=1,\\
	&&f(I,\theta  )=\frac{1}{vt\pi ^{2}\sqrt{1-\frac{I}{2vt}\cos^{2}\frac{\theta}{2}  } \sqrt{1-\frac{I}{2vt}\sin^{2}\frac{\theta}{2}  }},
\end{eqnarray}
and $I\in(0,2vt] $,$\theta\in [0,\pi ]$.

\section{ yield and error rate}
According to Eq. (\ref{PH}) and Eq. (\ref{PV}), we can calculate the theoretical value of $\langle Y_{n}  \rangle_{S_{K}^{i}}$ as
\begin{eqnarray}
	Y_{n}=&&\sum_{m=0}^{n} C_{n}^{m} (1-\eta^{chan})^{n-m}(\eta^{chan})^{m}(P_{k}^{m} +P_{l}^{m})      \\ \nonumber
	=&&(1-Pd)(1-\eta^{chan}\eta_{D} |f(l)|^{2})^{n}+\\\nonumber
	&&(1-Pd)(1-\eta^{chan}\eta_{D}|f(k)|^{2} )^{n}-\\\nonumber
	&&2(1-Pd)^{2}(1-\eta^{chan}\eta_{D}|f(l)|^{2} -\eta^{chan}\eta_{D}|f(k)|^{2})^{n}
	\\\nonumber
	=&&(1-Pd)(1-\eta^{chan}\eta_{D} |f(l)|^{2})^{n}+\\\nonumber
	&&(1-Pd)(1-\eta^{chan}\eta_{D}|f(k)|^{2} )^{n}-\\\nonumber
	&& 2(1-Pd)^{2}(1-\eta^{chan}\eta_{D})^{n},\\
	\langle Y_{n} \rangle_{S_{K}^{i}} =&&\frac{1}{	\langle P\rangle}_{S_{K}^{i}}\iiint_{S_{K}^{i}}Y_{n}f(I,\theta ,\phi )dI d\theta d\phi.
\end{eqnarray}

Similarly, taking state $ | k  \rangle  $ as an example,  the error rate can be calculated as
\begin{eqnarray}
	e_{n}Y_{n}=&&\sum_{m=0}^{n} C_{n}^{m}[e_{d}P_{k}^{m} +(1-e_{d})P_{l}^{m}] \label{eY}  \\\nonumber
	=&&e_{d}(1-Pd)(1-\eta^{chan}\eta_{D} |f(l)|^{2})^{n}+\\\nonumber
	&&(1-e_{d})(1-Pd)(1-\eta^{chan}\eta_{D}|f(k)|^{2} )^{n}-\\\nonumber
	&&(1-Pd)^{2}(1-\eta^{chan}\eta_{D})^{n},\\
	\langle e_{n} Y_{n} \rangle_{S_{K}^{i}} =&&\frac{1}{	\langle P\rangle_{S_{K}^{i}}}\iiint_{S_{K}^{i}}e_{n}Y_{n}f(I,\theta ,\phi )dI d\theta d\phi.
\end{eqnarray}

\section{The density matrix of output states} \label{rou}

For simplicity, we set $\bigtriangleup \theta_{X}=\bigtriangleup \theta_{Y}=\bigtriangleup \phi_{X}=\bigtriangleup \phi_{Y}=\bigtriangleup \theta_{Z}=\bigtriangleup$.
After post-selection, Bob's output $n$-photon states consist of three components: the $X$ basis, $Y$ basis, and $Z$ basis.

The $n$-photon states in the $Z$ basis can be derived as
\begin{eqnarray}
	\rho_{n,Z}=&&\int_{0 }^{I_{s}}dI \int_{ 0}^{2\pi }d\phi 
	(\int_{0 }^{\bigtriangleup }d\theta +\int_{ \pi-\bigtriangleup}^{\pi}d\theta )\\ \nonumber 
	&&|n\rangle_{a^{\dagger}(\theta,\phi)}\langle n| f(I,\theta )\frac{1}{2 \pi}\\ \nonumber 
	=&&\int_{0 }^{I_{s}}dI \int_{ 0}^{2\pi }d\phi 
	(\int_{0 }^{\bigtriangleup }d\theta +\int_{ \pi-\bigtriangleup}^{\pi}d\theta )\\ \nonumber 
	&&\frac{1}{n! }\sum_{m_{1}=0}^{n} \sum_{m_{2}=0}^{n}C_{n}^{m_{1}} C_{n}^{m_{2}}  \\ \nonumber 
	&&
	(\cos\frac{\theta}{2})^{m_{1}+m_{2}}(\sin\frac{\theta}{2})^{2n-m_{1}-m_{2}} \\ \nonumber 
	&&\sqrt{m_{1}!(n-m_{1})!}\sqrt{m_{2}!(n-m_{2})!} e^{-i\phi(m_{1}-m_{2})}  \\ \nonumber 
	&&f(I,\theta )\frac{1}{2 \pi}
	| m_{1} \rangle_{H}| n-m_{1} \rangle_{V}\langle m_{2}|_{H}\langle n-m_{2} |_{V}\\ \nonumber 
	=&&\int_{0 }^{I_{s}}dI  
	(\int_{0 }^{\bigtriangleup }d\theta +\int_{ \pi-\bigtriangleup}^{\pi}d\theta )\\ \nonumber 
	&&\sum_{m_{1}=0}^{n} C_{n}^{m_{1}}  
	(\cos\frac{\theta}{2})^{2m_{1}}(\sin\frac{\theta}{2})^{2n-2m_{1}} 
	f(I,\theta ) \\ \nonumber 
	&&
	| m_{1} \rangle_{H}| n-m_{1} \rangle_{V}\langle m_{1}|_{H}\langle n-m_{1} |_{V}\\ \nonumber 
	=&& \int_{0 }^{I_{s}}dI  
	\int_{0 }^{\bigtriangleup }d\theta \\ \nonumber 
	&&\sum_{m_{1}=0}^{n} C_{n}^{m_{1}}  ( (\cos\frac{\theta}{2})^{2m_{1}}(\sin\frac{\theta}{2})^{2n-2m_{1}} \\ \nonumber 
	&&+
	(\sin\frac{\theta}{2})^{2m_{1}}(\cos\frac{\theta}{2})^{2n-2m_{1}}   )f(I,\theta ) \\ \nonumber 
	&&
	| m_{1} \rangle_{H}| n-m_{1} \rangle_{V}\langle m_{1}|_{H}\langle n-m_{1} |_{V}\\ \nonumber 
\end{eqnarray}

For the single-photon states, the density matrix will be
\begin{eqnarray}
	\rho_{1,Z}=&& \int_{0 }^{I_{s}}dI  
	\int_{0 }^{\bigtriangleup }d\theta \sum_{m_{1}=0}^{1}  ( (\cos\frac{\theta}{2})^{2m_{1}}(\sin\frac{\theta}{2})^{2-2m_{1}} \nonumber \\ \nonumber 
	&&+
	(\sin\frac{\theta}{2})^{2m_{1}}(\cos\frac{\theta}{2})^{2-2m_{1}}   )f(I,\theta ) \\ \nonumber 
	&&
	| m_{1} \rangle_{H}| 1-m_{1} \rangle_{V}\langle m_{1}|_{H}\langle 1-m_{1} |_{V}\\ 
	=&&\int_{0 }^{I_{s}}dI  
	\int_{0 }^{\bigtriangleup }d\theta f(I,\theta )(| 1 \rangle_{V}\langle 1|_{V}+| 1 \rangle_{H}\langle 1|_{H})
	\quad 
\end{eqnarray}

For the two-photon states, the density matrix will be
\begin{eqnarray}
	\rho_{2,Z}=&& \int_{0 }^{I_{s}}dI  
	\int_{0 }^{\bigtriangleup }d\theta \sum_{m_{1}=0}^{2} C_{2}^{m_{1}} ( (\cos\frac{\theta}{2})^{2m_{1}}(\sin\frac{\theta}{2})^{4-2m_{1}}\nonumber  \\ \nonumber 
	&&+
	(\sin\frac{\theta}{2})^{2m_{1}}(\cos\frac{\theta}{2})^{4-2m_{1}}   )f(I,\theta ) \\ \nonumber 
	&&
	| m_{1} \rangle_{H}| 2-m_{1} \rangle_{V}\langle m_{1}|_{H}\langle 2-m_{1} |_{V}\\ \nonumber 
	=&&\int_{0 }^{I_{s}}dI  
	\int_{0 }^{\bigtriangleup }d\theta f(I,\theta )(\frac{1+\cos ^{2}\theta }{2} 
	(| HH \rangle\langle HH|\\ 
	&&+| VV \rangle\langle VV|)+\sin ^{2}\theta | HV \rangle\langle HV|)
\end{eqnarray}

The $n$-photon states in the $X$ basis can be derived as
\begin{eqnarray}
	\rho_{n,X}=&&\int_{0 }^{I_{s}}dI (\int_{ -\bigtriangleup}^{-\bigtriangleup }d\phi 
	+\int_{ \pi -\bigtriangleup}^{\pi -\bigtriangleup }d\phi)
	\int_{\frac{\pi}{2} -\bigtriangleup  }^{\frac{\pi}{2} +\bigtriangleup }d\theta\\ \nonumber 
	&&|n\rangle_{a^{\dagger}(\theta,\phi)}\langle n| f(I,\theta )\frac{1}{4 \bigtriangleup}\\ \nonumber 
	=&&\int_{0 }^{I_{s}}dI (\int_{ -\bigtriangleup}^{-\bigtriangleup }d\phi 
	+\int_{ \pi -\bigtriangleup}^{\pi -\bigtriangleup }d\phi)
	\int_{\frac{\pi}{2} -\bigtriangleup  }^{\frac{\pi}{2} +\bigtriangleup }d\theta\\ \nonumber 
	&&\frac{1}{n! }\sum_{m_{1}=0}^{n} \sum_{m_{2}=0}^{n}C_{n}^{m_{1}} C_{n}^{m_{2}}  \\ \nonumber 
	&&
	(\cos\frac{\theta}{2})^{m_{1}+m_{2}}(\sin\frac{\theta}{2})^{2n-m_{1}-m_{2}} \\ \nonumber 
	&&\sqrt{m_{1}!(n-m_{1})!}\sqrt{m_{2}!(n-m_{2})!} e^{-i\phi(m_{1}-m_{2})}  \\ \nonumber 
	&&f(I,\theta )\frac{1}{4 \bigtriangleup}
	| m_{1} \rangle_{H}| n-m_{1} \rangle_{V}\langle m_{2}|_{H}\langle n-m_{2} |_{V}\\ \nonumber 
	=&&\int_{0 }^{I_{s}}dI
	\int_{\frac{\pi}{2} -\bigtriangleup  }^{\frac{\pi}{2} +\bigtriangleup }d\theta\\ \nonumber 
	&&\sum_{m_{1}=0}^{n} C_{n}^{m_{1}}  
	(\cos\frac{\theta}{2})^{2m_{1}}(\sin\frac{\theta}{2})^{2n-2m_{1}} 
	f(I,\theta ) \\ \nonumber 
	&&
	| m_{1} \rangle_{H}| n-m_{1} \rangle_{V}\langle m_{1}|_{H}\langle n-m_{1} |_{V}\\ \nonumber 
	=&&\int_{0 }^{I_{s}}dI
	\int_{-\bigtriangleup  }^{\bigtriangleup }d\theta\\ \nonumber 
	&&\sum_{m_{1}=0}^{n} C_{n}^{m_{1}}  
	(\frac{1+\sin\theta}{2})^{m_{1}} (\frac{1-\sin\theta}{2})^{2n-2m_{1}} 
	 \\ \nonumber 
	&&f(I,\frac{\pi}{2}-\theta )
	| m_{1} \rangle_{H}| n-m_{1} \rangle_{V}\langle m_{2}|_{H}\langle n-m_{2} |_{V}\\ \nonumber 
	=&&\int_{0 }^{I_{s}}dI \int_{0  }^{\bigtriangleup }d\theta \\ \nonumber 
	&&\sum_{m_{1}=0}^{n} C_{n}^{m_{1}}  ((\frac{1+\sin\theta}{2})^{m_{1}} (\frac{1-\sin\theta}{2})^{n-m_{1}} \\ \nonumber 
	&&
	+(\frac{1-\sin\theta}{2})^{m_{1}} (\frac{1+\sin\theta}{2})^{n-m_{1}} )f(I,\frac{\pi}{2}-\theta )\\ \nonumber 
	&&
	| m_{1} \rangle_{H}| n-m_{1} \rangle_{V}\langle m_{2}|_{H}\langle n-m_{2} |_{V}
\end{eqnarray}
Notice that
\begin{eqnarray}
	f(I,\frac{\pi}{2}-\theta  )&&  =  \frac{1}{vt\pi ^{2}\sqrt{1-\frac{I}{2vt}\frac{1-\sin\theta}{2}  } 
		\sqrt{1-\frac{I}{2vt}\frac{1+\sin\theta}{2}}}  \nonumber \\
	&&=f(I,\frac{\theta-\pi}{2}  ).
\end{eqnarray}

For the single-photon states, the density matrix will be
\begin{eqnarray}
	\rho_{1,X}=&& \int_{0 }^{I_{s}}dI  
	\int_{0 }^{\bigtriangleup }d\theta \sum_{m_{1}=0}^{1}  ( (\frac{1+\sin\theta}{2})^{m_{1}}(\frac{1-\sin\theta}{2})^{1-m_{1}}\nonumber  \\ \nonumber 
	&&+
	(\frac{1-\sin\theta}{2})^{m_{1}}(\frac{1+\sin\theta}{2})^{1-m_{1}}   )f(I,\frac{\pi}{2}-\theta ) \\ \nonumber 
	&&
	| m_{1} \rangle_{H}| 1-m_{1} \rangle_{V}\langle m_{1}|_{H}\langle 1-m_{1} |_{V}\\ 
	=&&\int_{0 }^{I_{s}}dI  
	\int_{0 }^{\bigtriangleup }d\theta f(I,\frac{\pi}{2}-\theta )(| 1 \rangle_{V}\langle 1|+| 1 \rangle_{H}\langle 1|) \quad
\end{eqnarray}

For the two-photon states, the density matrix will be
\begin{eqnarray}
	\rho_{2,X}=&& \int_{0 }^{I_{s}}dI  
	\int_{0 }^{\bigtriangleup }d\theta \sum_{m_{1}=0}^{2} C_{2}^{m_{1}} ( (\frac{1+\sin\theta}{2})^{m_{1}}(\frac{1-\sin\theta}{2})^{2-m_{1}} \nonumber \\ \nonumber 
	&&+
	(\frac{1-\sin\theta}{2})^{m_{1}}(\frac{1+\sin\theta}{2})^{2-m_{1}}   )f(I,\frac{\pi}{2}-\theta ) \\ \nonumber 
	&&
	| m_{1} \rangle_{H}| 2-m_{1} \rangle_{V}\langle m_{1}|_{H}\langle 2-m_{1} |_{V}\\ 
	=&&\int_{0 }^{I_{s}}dI  
	\int_{0 }^{\bigtriangleup }d\theta f(I,\frac{\pi}{2}-\theta )(\frac{1+\sin ^{2}\theta }{2} 
	(| HH \rangle\langle HH|\nonumber \\ 
	&&+| VV \rangle\langle VV|)+\cos ^{2}\theta | HV \rangle\langle HV|)
\end{eqnarray}
Due to symmetry, the output states of the Y basis and the X basis are identical, and we have
\begin{eqnarray}
	\rho_{1,X}=\rho_{1,Y},\rho_{2,X}=\rho_{2,Y}.
\end{eqnarray}

After normalization, the density matrices for the single-photon and two-photon states output by Bob are
\begin{eqnarray}
	\rho_{1}&&=\rho_{1,X}+\rho_{1,Y}+\rho_{1,Z}=\frac{1}{2}(| H \rangle\langle H|+| V \rangle\langle V|), \\
	\rho_{2}&&=\rho_{2,X}+\rho_{2,Y}+\rho_{2,Z}\\ \nonumber 
	&&=(\int_{0 }^{I_{s}}dI  
	\int_{0 }^{\bigtriangleup }d\theta f(I,\theta )(\frac{1+\cos ^{2}\theta }{2} 
	(| HH \rangle\langle HH|\\ \nonumber 
	&&+| VV \rangle\langle VV|)+\sin ^{2}\theta | HV \rangle\langle HV|)\\ \nonumber 
	&&
	+2\int_{0 }^{I_{s}}dI  
	\int_{0 }^{\bigtriangleup }d\theta f(I,\frac{\pi}{2}-\theta )(\frac{1+\sin ^{2}\theta }{2} 
	(| HH \rangle\langle HH|\\ \nonumber 
	&&+| VV \rangle\langle VV|)+\cos ^{2}\theta | HV \rangle\langle HV|))/\\ \nonumber
	&&( \iint_{\sum S_{K}^{i} }  f(I,\theta ) dI d\theta ),
\end{eqnarray}

\section*{Acknowledgement}
 This work is supported by the National Natural Science Foundation of China under Grants  No. 12175106 and  No. 92365110, and the Postgraduate Research $\&$  Practice Innovation Program of Jiangsu Province under Grant No. KYCX25-1129.


\nocite{*}

	\begin{thebibliography}{99}
	
	

\bibitem{QSDC1}  G. L. Long, and  X. S. Liu, Theoretically efficient high-capacity quantum-key-distribution scheme,  Phys. Rev. A \textbf{65}, 032302 (2002).


\bibitem{QSDC2}  F. G. Deng, G. L. Long, and  X. S. Liu, Two-step quantum direct communication protocol using the Einstein-Podolsky-Rosen pair block,  Phys. Rev. A \textbf{68}, 042317 (2003).


\bibitem{QSDC3} F. G. Deng, and G. L. Long, Secure direct communication with a quantum one-time pad, Phys. Rev. A  \textbf{69}, 052319 (2004).


\bibitem{QSDChd} C. Wang, F. G. Deng, Y. S. Li, X. S. Liu, and G. L. Long, Quantum secure direct communication with high-dimension quantum superdense coding, Phys. Rev. A \textbf{71}, 044305 (2005).


\bibitem{li}  T. Li, and  G. L. Long, Quantum secure direct communication based on single-photon Bell-state measurement, New J. Phys. \textbf{22}, 063017 (2020).


\bibitem{Masking} G. L. Long, and H. Zhang, Drastic increase of channel capacity in quantum secure direct communication using masking, Sci. Bull. \textbf{66}, 1267-1269 (2021).


\bibitem{QSDCos} Y. B. Sheng, L. Zhou, and G. L. Long,  One-step quantum secure direct communication, Sci. Bull. \textbf{67}, 367-374 (2022).

\bibitem{QSDCdense} J. W. Wu, G. L. Long, and  M. Hayashi, Quantum secure direct communication with private dense coding using a general preshared quantum state, Phys. Rev. Appl. \textbf{17}, 064011 (2022).


\bibitem{QSDChy} Y. X. Xiao, L. Zhou, W. Zhong, M. M. Du, and Y. B. Sheng, The hyperentanglement-based quantum secure direct communication protocol with single-photon measurement, Quant. Inform. Process. \textbf{22}, 339 (2023).

\bibitem{QSDC22} Q. Zhang, M. M. Du, W. Zhong, Y. B. Sheng, and L. Zhou, Single-photon based three-party quantum secure direct communication with identity authentication,  Ann. Phys. (Berlin, Ger.) \textbf{536}, 3 (2024).



\bibitem{QSDC23} D. Pan,  G. L. Long, L. G. Yin,  Y. B  Sheng, D. Ruan, S. X. Ng,   J. H. Lu, and  L. Hanzo, The Evolution of Quantum Secure Direct Communication: On the Road to the Qinternet, IEEE Commun. Surv. Tutor. \textbf{26}, 1898-1949  (2024).


\bibitem{QSDCpath} G. L. Long, D. Pan, Y. B. Sheng, Q. Xue, J. Lu, and L. Hanzo, An evolutionary pathway for the quantum internet relying on secure classical repeaters, IEEE Netw. \textbf{36}, 82-88 (2022).

\bibitem{RDIQSDC}C. Liu, C. Zhang, S. P. Gu, X. F. Wang, L. Zhou, and Y. B. Sheng, Receiver-device-independent quantum secure direct communication, Sci. China: Phys. Mech. \& Astron. \textbf{68}, 5 (2025).

\bibitem{QSDC9} J. Y. Hu, B. Yu, M. Y. Jing, L. T. Xiao, S. T. Jia, G. Q. Qin, and G. L. Long, Experimental quantum secure direct communication with single photons,  Light: Sci. \& Appl. \textbf{5}, e16144 (2016).


\bibitem{QSDC10} W. Zhang, D. S. Ding, Y. B. Sheng, L. Zhou, B. S. Shi, and G. C. Guo, Quantum secure direct communication with quantum memory,  Phys. Rev. Lett. \textbf{118}, 220501	(2017).	

\bibitem{QSDC11} F. Zhu, W. Zhang,  Y. B. Sheng, and Y. D. Huang, Experimental long-distance quantum secure direct communication, Sci. Bull. \textbf{10}, 1519-1524 (2017).




\bibitem{decoy3}  D. Pan, Z. Lin, J. Wu, H. Zhang, Z. Sun, D. Ruan, L. Yin, G. L. Long, Experimental free-space quantum secure direct communication and its security analysis, Photonics Research \textbf{8}, 9 (2020).

\bibitem{Mapping} Z. W. Cao, L. Wang, K. X. Liang, G. Chai, and J. Y. Peng, Continuous-variable quantum secure direct communication based on Gaussian mapping, Phys. Rev. Appl. \textbf{16} 024012 (2021).

\bibitem{QSDC12}  Z. T. Qi, Y. H. Li, W. Y. Huang, J. Feng, Y. L. Zheng, and X. F. Chen,  A 15-user quantum secure direct communication network, Light: Sci. \& Appl. \textbf{10}, 183 (2021).

\bibitem{QSDC15}  H. R. Zhang, Z. Sun,  R. Y. Qi,  L. G. Yin,  G. L. Long, and  J. H. Lu,  Realization of quantum secure direct communication over 100 km fiber with time-bin and phase quantum states, Light: Sci \& Appl. \textbf{11}, 83 (2022).




\bibitem{QSDC20n} Z. W. Cao, Y. Lu, G. Chai, H. Yu, K. X. Liang, and L. Wang, Realization of quantum secure direct communication with continuous variable, Research \textbf{6}, 0193 (2023).

\bibitem{cvexp} I. Paparelle, F. Mousavi, F. Scazza, A. Bassi, M. Paris, and A. Zavatta, Practical quantum secure direct communication with squeezed states, 	arXiv:2306.14322 (2023).

\bibitem{QSDCpath24} D. Pan, Y. C. Liu, P. H. Niu, H. R. Zhang, F. H. Zhang, M. Wang, X. T. Song, X. W. Chen, C. Zheng, and G. L. Long, Simultaneous transmission of information and keyexchange using the same photonic quantum states, 	Sci. Adv. \textbf{11}, 8 (2024).

\bibitem{cvrecent}L. Wang, G. Chai, Z. Cao, X. Chen, K. Liang, and J. Peng, Quantum purification for coherent states and its application, Sci. China Phys. Mech. Astron. \textbf{68}, 220313 (2025).

\bibitem{QSDCnet}	Y. L. Yang, Y. H. Li, H. Li, C. N. Wu, Y. L. Zheng, and X. F. Chen, A 300-km fully-connected quantum secure direct communication network, Sci. Bull. \textbf{70}, 1445 (2025).








\bibitem {PQKD1} E. Diamanti, H. K. Lo, B. Qi, and  Z. L. Yuan,   Practical challenges in quantum key distribution, npj Quantum Information \textbf{2},  16025 (2016).

\bibitem{qkd3} F. H. Xu, X. F. Ma, Q. Zhang, H. K. Lo, and J. W. Pan, Secure quantum key distribution with realistic devices, Rev. Mod. Phys. \textbf{92}, 025002 (2020).

\bibitem {QKDsp}  M. Bozzio, M. Vyvlecka, M. Cosacchi, C. Nawrath, T. Seidelmann, J. C. Loredo, S. L. Portalupi, V. M. Axt, P. Michler, and P. Walther,     Enhancing quantum cryptography with quantum dot single-photon sources, npj Quantum Information \textbf{8}, 104 (2022).




\bibitem{collective2} J. Wu, Z. Lin, L. Yin, and G. L. Long, Security of quantum secure direct communication based on Wyner's wiretap channel theory,	 Quantum Eng. \textbf{1}, e26 (2019).


\bibitem {PQSDC1} R. Y. Qi, Z. Sun, Z. S. Lin, P. H. Niu, W. T. Hao, L. Y. Song, Q. Huang, J. C. Gao, L. G. Yin, and  G. L. Long,   Implementation and security analysis of practical quantum secure direct communication,  Light: Sci. Appl.  \textbf{8},  22 (2019).


\bibitem{decoy4}  X. Liu, Z. J. Li, D. Luo, C. F. Huang, D. Ma, M. M. Geng, J. W. Wang, Z. R. Zhang, and K. J. Wei, Practical decoy-state quantum secure direct communication, Sci. China: Phys. Mech. \& Astron. \textbf{64}, 120311 (2021).










\bibitem{QHACK1} G. Brassard, N. L\"utkenhaus, T. Mor, and B. C. Sanders,   Limitations on Practical Quantum Cryptography,  Phys. Rev. Lett. \textbf{85}, 1330 (2000).
\bibitem{QHACK2}  L. Lydersen, C. Wiechers, C. Wittmann, D. Elser, J. Skaar, and  V. Makarov,    Hacking commercial quantum cryptography systems by tailored bright illumination,   Nat. Photonics \textbf{4}, 686-689 (2010).
\bibitem{QHACK3}   I. Gerhardt, Q. Liu, A. L. Linares, J.  Skaar, C. Kurtsiefer, and  V. Makarov,  Full-field implementation of a perfect eavesdropper on a quantum cryptography system,  	Nat. Commun.  \textbf{2}, 349 (2011).
\bibitem{QHACK4}   S. H. Sun, M. Gao, M. S. Jiang, C. Y. Li, and L. M. Liang, Partially random phase attack to the practical two-way quantum-key-distribution system,  Phys. Rev. A \textbf{85}, 032304  (2012).
\bibitem{QHACK5} S. H. Sun, F. H. Xu, M. S. Jiang, X. C. Ma, H. K. Lo, and L. M. Liang,    Effect of source tampering in the security of quantum cryptography,  Phys. Rev. A \textbf{92}, 022304  (2015).


\bibitem{qkd2} H. K. Lo, M. Curty, and B. Qi, Measurement-device-independent quantum key distribution, Phys. Rev. Lett. \textbf{108}, 130503 (2012).

\bibitem {MDIQKD2}  Y. Liu, T. Y. Chen, L. J. Wang, H. Liang, G. L. Shen-tu, J. Wang, K. Cui, H. L. Yin, N. L. Liu, L.  Li, X. F. Ma, J. S. Pelc, M. M. Fejer, C. Z. Peng, Q. Zhang, and J. W. Pan,   Experimental Measurement-Device-Independent Quantum Key Distribution,  Phys. Rev. Lett. \textbf{111}, 130502 (2013).

\bibitem {MDIQKD3}  H. L. Yin, T. Y. Chen, Z. W. Yu, H. Liu, L. X.  You, Y. H. Zhou, S. J.  Chen, Y. Q. Mao, M. Q. Huang, W. J. Zhang, H. Chen, M. J. Li, D. Nolan, F. Zhou, X. Jiang, Z. Wang, Q. Zhang, X. B. Wang, and J. W. Pan,  Measurement-Device-Independent Quantum Key Distribution Over a 404 km Optical Fiber,  Phys. Rev. Lett. \textbf{117}, 190501 (2016).
\bibitem {MDIQKD4}  Y. Cao, Y. H. Li, K. X. Yang, Y. F. Jiang, S. L. Li, X. L. Hu, M. Abulizi, C. L. Li, W. J. Zhang, Q. C. Sun, W. Y. Liu, X. Jiang, S. K. Liao, J. G. Ren, H. Li, L. X. You, Z.  Wang, J. Yin, C. Y. Lu, X. B. Wang, Q. Zhang, C. Z. Peng, and J. W. Pan,  Long-Distance Free-Space Measurement-Device-Independent Quantum Key Distribution,  Phys. Rev. Lett. \textbf{125}, 260503 (2020).
\bibitem {MDIQKD1} D. Luo, X. Liu, K. B. Qin, Z. R. Zhang, and K. J. Wei,   Practical asynchronous measurement-device-independent quantum key distribution with advantage distillation, Phys. Rev. A \textbf{110}, 022605  (2024).





\bibitem{QSDC14n} P. H. Niu, Z. R. Zhou, Z. S. Lin, Y. B. Sheng, L. G. Yin, and G. L. Long, Measurement-device-independent quantum communication without encryption, Sci. Bull. \textbf{63}, 1345-1350 (2018).


\bibitem{MDI2} Z. R. Zhou, Y. B. Sheng, P. H. Niu, L. G. Yin, G. L. Long, and L. Hanzo, Measurement-device-independent quantum secure direct communication, Sci. China: Phys. Mech. \& Astron. \textbf{63}, 230362 (2020).




\bibitem{MDI3} X. D. Wu, L. Zhou, W. Zhong, Y. B. Sheng, High-capacity measurement-device-independent quantum secure direct communication, Quant. Inf. Process. \textbf{19} 10354 (2020).

\bibitem{QSDC16n} J. W. Ying, L. Zhou, W. Zhong, and Y. B. Sheng, Measurement-device-independent one-step quantum secure direct communication, Chin. Phys. B \textbf{31}, 120303 (2022).

\bibitem{QSDC19} Y. P. Hong, L. Zhou, W. Zhong, and Y. B. Sheng, Measurement-device-independent three-party quantum secure direct communication, Quant. Inform. Process. \textbf{22}, 111 (2023).

\bibitem{QSDC24}J. Liu, X. Zou, X. Wang, Y. Chen, Z. Rong, Z. Huang, S. Zheng, X. Liang, and J. Wu, Applying a class of general maximally entangled states in measurement-device-independent quantum secure direct communication, Phys. Rev. Appl. \textbf{21}, 044010 (2024).



\bibitem{qkd1} A. Ac\'{\i}n, N. Brunner, N. Gisin, S. Massar, S. Pironio, and V. Scarani, Device-independent security of quantum cryptography against collective attacks, Phys. Rev. Lett. \textbf{98}, 230501 (2007).

\bibitem{DI}  L. Zhou,  Y. B. Sheng, and G. L. Long, Device-independent quantum secure direct communication against collective attacks, Sci. Bull.  \textbf{65}, 12-20 (2020).

\bibitem {DIQKD} S. Pironio, A. Ac\'{\i}n, N. Brunner, N. Gisin, S. Massar and V. Scarani, Device-independent quantum key distribution secure against collective attacks, New J. Phys. \textbf{11}, 045021 (2009).
\bibitem{DIQKD1}  J. Ko{\l}ody\'{N}ski, A. M\'{a}ttar, P. Skrzypczyk, E. Woodhead, D. Cavalcanti, K. Banaszek, and A. Ac\'{\i}n, Device-independent quantum key distribution with single-photon sources, Quantum \textbf{4}, 260 (2020).


\bibitem{QSDC16}  L. Zhou,  and  Y. B. Sheng, One-step device-independent quantum secure direct communication, Sci. China: Phys. Mech. \& Astron. \textbf{65}, 250311 (2022).

\bibitem{QSDC18} L. Zhou, B. W. Xu, W. Zhong, and Y. B. Sheng, Device-independent quantum secure direct communication with single-photon sources, Phys. Rev. Appl. \textbf{19}, 014036 (2023).	
\bibitem{QSDC21} H. Zeng, M. M. Du, W. Zhong, L. Zhou, and Y. B. Sheng, High-capacity device-independent quantum secure direct communication based on hyper-encoding, Funda. Res. \textbf{4}, 852-858 (2024).


\bibitem{DIE1} D. P. Nadlinger, P. Drmota, B. C. Nichol, G. Araneda, D. Main, R. Srinivas, D. M. Lucas, C. J. Ballance, K. Ivanov, and E. Z. Tan, Experimental quantum key distribution certified by Bell's theorem, Nature \textbf{607}, 682-686 (2022).
\bibitem{DIE2} W. Zhang, T. V. Leent, K. Redeker, R. Garthoff, R. Schwonnek, F. Fertig, S. Eppelt, W. Rosenfeld, V. Scarani, and C. C. W. Lim, A device-independent quantum key distribution system for distant users, Nature \textbf{607}, 687-691 (2022).
\bibitem{DIE3} W. Z. Liu, Y. Z. Zhang, Y. Z. Zhen, M. H. Li, Y. Liu, J. Y. Fan, F. H. Xu, Q. Zhang, and J. W. Pan, Toward a photonic demonstration of device-independent quantum key distribution, Phys. Rev. Lett. \textbf{129}, 050502 (2022).

\bibitem{DIreview} I. W. Primaatmaja, K. T. Goh, E. Y.-Z. Tan, J. T.-F. Khoo, S. Ghorai, and C. Lim, Security of device-independent quantum key distribution protocols: a review, Quantum, \textbf{7}, 932 (2023).


\bibitem{decoy} H.K. Lo, X.F. Ma, and K. Chen,  Decoy state quantum key distribution, Phys. Rev. Lett. \textbf{94}, 230504 (2005).
\bibitem{decoy1} X.F. Ma, B. Qi, Y. Zhao, and H.K. Lo, Practical decoy state for quantum key distribution, Phys. Rev. A \textbf{72}, 012326 (2005).

	\bibitem{sidechannel} A. Gnanapandithan, L. Qian, and H. K. Lo, Hidden Multidimensional Modulation Side Channels in Quantum Protocols,  Phys. Rev. Lett. \textbf{134}, 130802 (2025).

\bibitem{trojan0} A. Vakhitov, V. Makarov, and D. R. Hjelme,  Large pulse attack as a method of conventional optical eavesdropping in quantum cryptography,   J. Mod. Opt. \textbf{48}, 2023 (2001).

\bibitem{trojan1} N. Gisin, S. Fasel, B. Kraus, H. Zbinden, and G. Ribordy, Trojan-horse attacks on quantum-key-distribution systems, Phys. Rev. A \textbf{73}, 022320 (2006).

\bibitem{trojan2}  \'A. Navarrete, and M. Curty,   Improved finite-key security analysis of quantum key distribution against Trojan-horse attacks, Quantum Science and Technology,  \textbf{7}, 3 (2022).

\bibitem{trojan3}  T. T. Luo, Q. Liu, X. R. Sun, C. F. Huang, Y. Chen, Z. R. Zhang, and K. J. Wei,   Security analysis against the Trojan horse attack on practical  polarization-encoding quantum key distribution systems,    Phys. Rev. A \textbf{109}, 042608 (2024).





\bibitem{passivedecoy1}  Q. Wang, C. H. Zhang, and X. B. Wang, Scheme for realizing passive quantum key distribution with heralded single-photon sources, Phys. Rev. A \textbf{93}, 032312 (2016).

\bibitem{passivedecoy2} C. H. Zhang, C. M. Zhang, and Q. Wang, Efficient passive measurement-device-independent quantum key distribution, Phys. Rev. A \textbf{99}, 052325 (2019).


\bibitem{passivedecoy3} M. Curty, T. Moroder, X. F. Ma, and N. L\"{u}tkenhaus, Non Poissonian statistics from Poissonian light sources with application to passive decoy state quantum key distribution, Opt. Lett. \textbf{34}, 3238-3240 (2009).

\bibitem{passivedecoy4} M. Curty, X. F. Ma, B Qi, and T Moroder, Passive decoy-state quantum key distribution with practical light sources, Phys. Rev. A \textbf{81},  022310 (2010).

\bibitem{passivedecoy5}Q. C. Sun, W. L. Wang, Y. Liu, F. Zhou, J. S. Pelc, M. M. Fejer, C. Z. Peng, X. F. Chen, X. F. Ma, Q Zhang, and J. W. Pan,   Experimental passive decoy-state quantum key distribution, 	Laser Phys. Lett. \textbf{11}, 085202 (2014).

\bibitem{passivedecoy6} S. H. Sun, G. Z. Tang, C. Y. Li, and L. M. Liang, Experimental demonstration of passive decoy-state quantum key distribution with two independent lasers, Phys. Rev. A \textbf{94}, 032324 (2016).

\bibitem{passivedecoy} J. W. Ying, P. Zhao, W. Zhong, M. M. Du, X. Y. Li, S. T. Shen, A. L. Zhang, L. Zhou, and Y. B. Sheng, Passive decoy-state quantum secure direct communication with heralded single-photon source, Phys. Rev. Appl. \textbf{22}, 024040 (2024).

\bibitem{passivestate1}  M. Curty, X. F. Ma, H. K. Lo, and N. L\"utkenhaus, Passive sources for the Bennett-Brassard 1984 quantum-key-distribution protocol with practical signals, Phys. Rev. A \textbf{82}, 052325 (2010).

\bibitem{passivestatecv1}  B. Qi, P. G. Evans, and W. P. Grice, Passive state preparation in the Gaussian-modulated coherent-states quantum key distribution,  Phys. Rev. A \textbf{97}, 012317 (2018).


\bibitem{passivestatecv2}  B. Qi, H. Gunther, P. G. Evans, B. P. Williams, R. M. Camacho, and N. A. Peters, Experimental Passive-State Preparation for Continuous-Variable Quantum Communications, Phys. Rev. Applied \textbf{13}, 054065 (2020).


\bibitem{passivestatecv3}  F. Y. Ji, P. Huang, T. Wang, X. Q. Jiang, and G. H. Zeng,  Gbps key rate passive-state-preparation continuous-variable quantum key distribution within an access-network area, Photonics Res. \textbf{12}, 1485 (2024).

\bibitem{QSDCpc} J. W. Ying, J. Y. Wang, Y. X. Xiao, S. P. Gu, X. F. Wang, W. Zhong, M. M. Du, X. Y. Li, S. T. Shen, A. L. Zhang, L. Zhou, and Y. B. Sheng, Passive-state preparation for quantum secure direct communication, Sci. China: Phys. Mech. \& Astron. \textbf{68}, 240312 (2025).

\bibitem{passive3} W. Y. Wang, R. Wang, C. Q. Hu, V. Zapatero, L. Qian, B. Qi, M. Curty, and H. K. Lo, Fully passive quantum key distribution, Phys. Rev. Lett. \textbf{130}, 220801 (2023).	



\bibitem{passive5} V. Zapatero, W. Wang, and M. Curty, A fully passive transmitter for decoy-state quantum key distribution, Quantum Sci. Technol. \textbf{8},  025014 (2023).

\bibitem{passive6} V. Zapatero, and M. Curty, Finite-key security of passive quantum key distribution, 	Phys. Rev. Appl. \textbf{21}, 014018 (2024).

\bibitem{passive7}J. J. Li, W. Y. Wang, and H. K. Lo, Fully passive measurement-device-independent quantum key distribution, 	Phys. Rev. Appl. \textbf{21}, 064056 (2024).


\bibitem{passive8}X. Wang, F. Y. Lu, Z. H. Wang, Z. Q. Yin, S. Wang, J. Q. Geng, W. Chen, D. Y. He, G. C.  Guo, and Z. F. Han, Fully passive measurement-device-independent quantum key distribution, 	Phys. Rev. Appl. \textbf{21}, 064067 (2024).

\bibitem{FPTW}   W. Y. Wang, R. Wang, and H. K. Lo, Fully-Passive Twin-Field Quantum Key Distribution, 	arXiv:2304.12062.

\bibitem{FPCKA} J. J. Li, W. Y. Wang, and H. F. Chau, Fully Passive Quantum Conference Key Agreement, 	arXiv:2407.15761.

\bibitem{passive9}  C. Q. Hu, W. Y. Wang, K. S. Chan, Z. H. Yuan, and H. K. Lo,  Proof-of-Principle Demonstration of Fully Passive Quantum Key Distribution,  Phys. Rev. Lett. \textbf{131}, 110801 (2023).

\bibitem{passive4} F. Y. Lu, Z. H. Wang, V. Zapatero, J. L. Chen, S. Wang, Z. Q. Yin, M. Curty, D. Y. He, R. Wang, W. Chen, G. J. Fan-Yuan, G. C. Guo, and Z. F. Han, Experimental demonstration of fully passive quantum key distribution, Phys. Rev. Lett. \textbf{131}, 110802 (2023).	

\bibitem{Wyner}	A. D. Wyner, The wire-tap channel,  Bell Syst. Tech. J. \textbf{54}, 1355-1387 (1975).
	\bibitem{Unambiguous} Y. Feng,  R. Y. Duan,  and M. S. Ying,   Unambiguous discrimination between mixed quantum states, Phys. Rev. A \textbf{70}, 012308 (2004).
	
	
	
	
	
	
	
	
	
	
	
	
	
	
	
	
	
	
	
	
	
	
	
	
	
	
\end{thebibliography}
\end{document}